\documentclass[%
showpacs,preprintnumbers,
 amsmath,amssymb,
 aps,
 prd,
 longbibliography,
floatfix,
 lengthcheck,%
]{revtex4}

\usepackage{graphicx}
\usepackage{hyperref}

\begin{document}

\title{The sensitivity of neutrinos to the supernova turbulence power spectrum: point source statistics}

\author{James P.\ Kneller}
\email{jpknelle@ncsu.edu}
\affiliation{Department of Physics, North Carolina State University, Raleigh, North Carolina 27695, USA}

\author{Neel V.\ Kabadi}
\email{nvkabadi@ncsu.edu}
\affiliation{Department of Physics, North Carolina State University, Raleigh, North Carolina 27695, USA}

\begin{abstract}
The neutrinos emitted from the proto-neutron star created in a core-collapse supernova must run through a significant amount of turbulence before exiting the star. 
Turbulence can modify the flavor evolution of the neutrinos imprinting itself upon the signal detected here at Earth. 
The turbulence effect upon individual neutrinos, and the correlation between pairs of neutrinos, might exhibit sensitivity to the power spectrum of the turbulence and recent analysis of 
the turbulence in a two-dimensional hydrodynamical simulation of a core-collapse supernova indicates the power spectrum may not be the Kolmogorov $5/3$ inverse power law 
as has been previously assumed. In this paper we study the effect of non-Kolmogorov turbulence power spectra upon neutrinos from a point source as a function 
of neutrino energy and turbulence amplitude at a fixed post-bounce epoch. We find the two effects of turbulence upon the neutrinos - the distorted phase effect and the 
stimulated transitions - both possess strong and weak limits in which dependence upon the power spectrum is absent or evident respectively. 
Since neutrinos of a given energy will exhibit these two effects at different epochs of the supernova each with evolving strength, we find there is sensitivity to the 
power spectrum present in the neutrino burst signal from a Galactic supernova. 
\end{abstract}

\date{\today}

\maketitle


\section{Introduction}

There is now ample evidence from both observations and numerical simulations for the multi-dimensional nature of core-collapse supernovae. 
The high-velocity ``jets'' of sulfur-rich material - which presumably originated deep in the stellar mantle - seen in the supernova remnant Cassiopeia A \cite{2013ApJ...772..134M},
the double-peaked structure of the Oxygen and Magnesium nebular lines in observations of SN 2003jd \cite{2005Sci...308.1284M}, and the
spectropolarimetric observations of stripped-envelope core-collapse supernovae \cite{2012ApJ...754...63T} can all be explained if the explosions were aspherical.
Asphericity in the hydrodynamical simulations of core collapse supernovae is seen to emerge even when the progenitor is spherically symmetric 
\cite{2011ApJ...742...74M,2012arXiv1210.5241D,2012arXiv1210.6674O,2012ApJ...755..138H,2012ApJ...746..106P,2012ApJ...761...72M,2012ApJ...749...98T,2013arXiv1301.1326L,2014ApJ...792...96T,2014ApJ...785..123C}. 
If the asphericity is indeed generated deep with the star during the earliest epochs of the explosion then one would naturally expect the generation of turbulence in the fluid. The 
turbulence, which some have argued is crucial for achieving an explosion \cite{2015ApJ...799....5C}, would, in turn, alter the flavor evolution of neutrinos racing through 
the stellar mantle from the cooling proto-neutron star formed at the core. 

Finding the consequence of the turbulence upon the neutrinos is vital for interpreting the signal from the next supernova in the Galaxy. This need has long been 
recognized and various authors have examined the effect of turbulence upon neutrinos
\cite{Loreti:1995ae,Fogli:2006xy,Friedland:2006ta,Choubey,2010PhRvD..82l3004K,Reid,Kneller:2012id,2013PhRvD..88b3008L,2014PhRvD..89g3022P,2015PhRvD..91b5001P}. 
From these studies it has emerged that turbulence can affect the neutrinos in two different ways. 
The first, more direct effect of the turbulence is to `stimulate' transitions between the instantaneous neutrino eigenstates \cite{2013PhRvD..88b3008L,2014PhRvD..89g3022P,2015PhRvD..91b5001P} while the neutrino is propagating 
through the turbulent region. Although this effect depends upon a number of factors, typically noticeable turbulence effects require the density fluctuations to be present in the region of the supernovae mantle 
where neutrinos experience the Mikheyev, Smirnov and Wolfenstein (MSW) resonances \cite{M&S1986,Wolfenstein1977} and their amplitude must be of order a few percent. 
That said, the description of the stimulated transition effect of turbulence is not in terms of MSW resonances and MSW resonances are not required for the effect to appear. 
The second, more subtle effect of turbulence occurs when the neutrino transition probabilities exhibit phase effects \cite{2006PhRvD..73e6003K,2007PhRvD..75i3002D}.
In order to observe phase effects and this second, indirect, sensitivity to turbulence we require at least two semi-adiabatic MSW resonances and/or density discontinuities in the profile. 
Even then, it is sometimes possible to reduce the imprint of this second effect by carefully selecting the profile and neutrino energy. 
In more general circumstances we find both effects simultaneously though the second effect of turbulence becomes most obvious 
when the amplitude is small because the direct effect is usually negligible in this limit \cite{2010PhRvD..82l3004K}.

While the basic effects of turbulence upon the neutrinos have been determined, it is not apparent to what extent they might operate in a supernova 
due to the lack of suitable three-dimensional, high resolution, long duration hydrodynamical simulations. In their absence authors have been forced to 
model the turbulence in a supernova by adopting a turbulence-free profile and then inserting turbulence into it in the form of a random field with 
assumed properties. The problem with this approach is that the validity of these prescriptions for the turbulence in supernovae are unknown. 
That situation changed recently with the study by Borriello \emph{et al.} \cite{2013arXiv1310.7488B} of the turbulence in a two-dimensional 
simulation from Kifonidis \emph{et al.} \cite{2006A&A...453..661K} which approached the necessary resolution and duration. 
Borriello \emph{et al.} fitted the power spectrum for the turbulence along each radial sliceof the simulation with a broken inverse-power law 
defined by four parameters. Two of these parameters correspond to spectral indices which they called $p_{1}$ and $p_2$: $p_1$ is the index for the longer wavelengths 
and the other, $p_2$, for the shorter. The other two parameters are the amplitude and the break wavenumber defined in terms of a multiple of the long wavelength cutoff scale.
The short wavelength index was found to have a mean and $1-\sigma$ error of around $p_2=3.04^{+0.57}_{-0.63}$ while the index for the longer wavelengths was found to have a mean and $1-\sigma$ error 
of $p_1=1.85^{+0.54}_{-0.77}$. 
The analysis by Borriello \emph{et al.} is a welcome addition to the literature but the well-known differences between the properties of 
turbulence in two and three spatial dimensions means it is not clear which results can be safely carried over to 3D. For example, in the inertial range of wavenumbers for 2D turbulence one may 
observe a Kraichnan inverse enstrophy cascade which funnels turbulent power into the long wavelength modes, a Kolmogorov energy cascade in the opposite direction, and even 
double cascades \cite{2006JLTP..145...25G} where turbulence is injected at some given scale and cascades to both longer \emph{and} shorter wavelengths. The turbulence seen 
by Borriello \emph{et al.} appears to be of this double cascade type because they find broken power laws similar to a Nastrom-Gage spectrum \cite{1985JAtS...42..950N}. If the presence of the `break wavenumber' and a short wavelength index $p_2$ is due to the 2D nature of the turbulence and if the amplitude they obtain is similarly contaminated by the inverse cascade effect, this leaves, perhaps, just the long wavelength index $p_1$ as being transferable to a 3D study. 
Thus, the most conservative conclusion to draw from the study is that the long wavelength spectral index has a mean of $p_1 \sim 5/3$ and the fact it has a range 
appears to indicate the turbulence is also not always `fully developed' justifying the exploration of something other 
than a Kolmogorov, $5/3$, power spectrum in the prescriptions for turbulence in 3D which has heretofore been the default. Changing the power spectrum will alter the evolution of 
individual neutrinos passing through the turbulence and the correlation between pairs of neutrinos sent along parallel rays \cite{2013PhRvD..88d5020K}.

The analytic results of Friedland \& Gruzinov \cite{Friedland:2006ta} and Patton, Kneller \& McLaughlin \cite{2014PhRvD..89g3022P,2015PhRvD..91b5001P} can be used to predict the 
effect of changing the spectral index for the direct, stimulated transition, effect of turbulence. They indicate that a lowering of the index (hardening) of the power spectrum should 
increase the stimulated transition effect upon the neutrinos by i) increasing the amplitude of the turbulence modes which lead to transitions between the neutrino states, ii) 
permitting more combinations of modes to drive transitions without a severe simultaneous narrowing of the resonance, and iii) lowering the amplitude of the modes which suppress those transitions. 
However the precise amount by which the direct turbulence effect alters the neutrino transition properties as the power spectrum changes 
has not been determined, and nothing exists in the literature for the indirect turbulence effect of distorted phase. It is the filling of these holes which is the goal 
of this paper. We begin by describing the calculations we have performed paying particular attention to the turbulence power spectrum we use and other details. 
The section following demonstrates the two effects of turbulence and the two descriptions which we shall use to make predictions and understand our 
results. Our results for the change in the ensembles for single neutrinos at three different energies and a wide range of 
turbulence amplitudes as a function of the power spectral index are then presented and we finish with a summary and our conclusions where we attempt to construct a coherent 
description of turbulence effects in supernova neutrinos. 


\section{Description of the calculations}

In order to study the effect of the supernova turbulence upon the neutrinos we compute the set of probabilities 
that a neutrino initially in some state $\nu_j$ is later detected in some other state $\nu_{i}$ at a different location - the transition probabilities. These quantities are denoted as $P_{ij}$ 
for neutrinos and $\bar{P}_{ij}$ for the antineutrinos. The transition probabilities can be computed from the elements of the evolution matrix $S$ which links 
the initial and final neutrino states, that is $P_{ij}=|S_{ij}|^2$, and the evolution matrix is computed by solving the Schrodinger equation
\begin{equation}
\imath \frac{dS}{d\lambda} = H\,S \label{eq:S}
\end{equation} 
where $H$ is the Hamiltonian and $\lambda$ the affine parameter along the neutrino trajectory. 
The transition probabilities we report in this paper are for the `matter' basis states. The matter basis states are related to the flavor basis states by the matter mixing matrix $\tilde{U}$ which 
is defined so that the flavor basis Hamiltonian $H^{(f)}$ and its eigenvalue matrix $\tilde{K}$ are related via $H^{(f)} = \tilde{U}\tilde{K}\tilde{U}^{\dagger}$ \cite{2009PhRvD..80e3002K,2012JPhG...39c5201G}. 
In our case the Hamiltonian possesses two contributions: the first, $H_0$, is from the vacuum and the second, $H_{MSW}$ comes from the effect of the matter upon the neutrino \cite{M&S1986,Wolfenstein1977}.
We do not include the contribution to $H$ from `collective' effects: see Duan, Fuller \& Qian \cite{2010ARNPS..60..569D} for a review of this fascinating subject. 
The vacuum Hamiltonian for a neutrino of a given energy $E$ is defined by the two independent mass squared differences $\delta m_{ij}^2 = m_i^2 - m_j^2$ of $\delta m_{32}^2$ and $\delta m_{21}^2$. 
It is diagonal in the `mass' basis which is related to the flavor basis by the Maki-Nakagawa-Sakata-Pontecorvo \cite{Maki:1962mu,PDG} unitary `mixing' matrix $U$. 
The mixing matrix can be written in terms of three vacuum mixing angles, $\theta_{12}$, $\theta_{13}$ and $\theta_{23}$, a CP phase $\delta$, and two Majorana phases though the two 
Majorana phases do not influence the evolution \cite{1987NuPhB.282..589L,2012JPhG...39c5201G}. 
Throughout this paper we adopt numerical values of $\delta m_{21}^2 = 7.5 \times 10^{-5}\;{\rm eV^2}$,  $\delta m_{32}^2 = 2.32 \times10^{-3}\;{\rm eV^2}$ (a normal hierarchy), 
$\theta_{12} =33.9^{\circ}$, $\theta_{13}=9^{\circ}$ and $\theta_{23}=45^{\circ}$ which are consistent with the values from the Particle Data Group \cite{PDG}. The CP phase is set to zero. 

The MSW potential $H_{MSW}$ is diagonal in the flavor basis because matter interacts with neutrinos based on their flavor. The matter affects the neutrinos via both neutral and charged current 
channels but the neutral current contribution to $H_{MSW}$ may be ignored because it leads to an unobservable global phase. In contrast, the absence of mu and tau leptons in the matter means 
the charged current potential affects the electron flavor neutrino and antineutrinos only. 
The charged current potential for the electron flavor neutrinos and antineutrinos is given by $V_{ee} =\pm \sqrt{2} G_F n_e(r)$ where $G_F$ is the Fermi constant and $n_e(r)$ the electron density.
The plus sign applies to the electron neutrinos, the minus sign for the electron antineutrinos. This potential is the `ee' element of $H_{MSW}$. 
The tiny radiative $\mu\tau$ potential \cite{1987PhRvD..35..896B,2008PhRvD..77f5024E} is ignored since it is a factor of $\sim 10^{-5}$ smaller than the potential 
affecting the electron flavor in the standard model (but may be two or three orders of magnitude bigger if supersymmetric contributions are included \cite{2010PhRvD..81a3003G}).
It is through the electron density $n_e(r)$ that the turbulence enters $H_{MSW}$. 
As noted in the Introduction, the ideal would be to use density profiles taken from high resolution, long duration, three-dimensional simulations of 
supernovae in order to study the effect of turbulence. These are not available so one is forced to adopt a profile from a one-dimensional simulation and add turbulence to it. 
We shall introduce the turbulence in such a way that the profile from the 1D simulation is also the mean electron density $\langle n_e(r)\rangle$, the average here being over 
realizations of the turbulence. 
\begin{figure}
\includegraphics[clip,width=\linewidth]{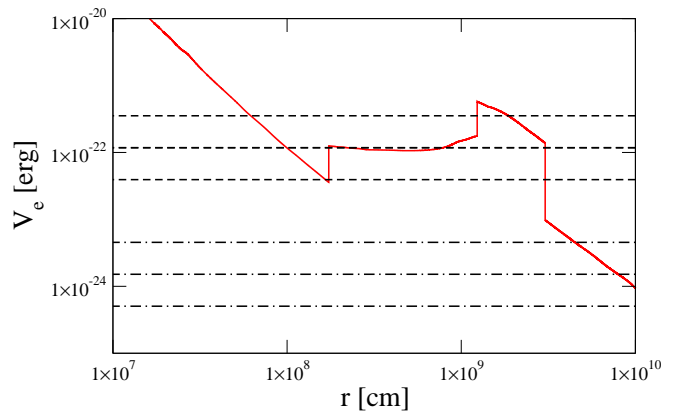}
\caption{The MSW potential for the snapshot at $t=3\;{\rm s}$ post-bounce of the $M=10.8\;M_{\odot}$ simulation from Fischer \emph{et al.}  \cite{Fischer:2009} as a function of distance. 
From the inside out, the three discontinuities in the profile are the reverse shock at $r_r =1,734\;{\rm km}$, the contact discontinuity at $r_c =12,348\;{\rm km}$, and 
the forward shock at $r_f =30,323\;{\rm km}$.
From top to bottom, the horizontal dashed lines are the two-flavor resonance potential for a neutrino with $E=5\;{\rm MeV}$, $E=15\;{\rm MeV}$, and $E=45\;{\rm MeV}$ respectively using 
mixing parameters $\delta m^2 = 2.32 \times10^{-3}\;{\rm eV^2}$ and $\theta=9^{\circ}$ while the the horizontal dot-dashed lines are the two-flavor resonance potential for a neutrino with $E=5\;{\rm MeV}$, $E=15\;{\rm MeV}$, and $E=45\;{\rm MeV}$ from top to bottom respectively using mixing parameters $\delta m^2 = 7.5 \times10^{-5}\;{\rm eV^2}$ and $\theta=33.9^{\circ}$. \label{fig:profile}}
\end{figure}
The profile we use for $\langle n_e(r)\rangle$  is the $t=3\;{\rm s}$ postbounce snapshot from the $10.8 \;M_{\odot}$ simulation by Fischer \emph{et al.} \cite{Fischer:2009}. 
The MSW potential for this profile is shown in figure (\ref{fig:profile}) and was chosen so that neutrinos with the MSW resonance of the three energies we shall use throughout this paper, $5\;{\rm MeV}$, $15\;{\rm MeV}$ and $45\;{\rm MeV}$, intersect the profile in the region where we shall place the turbulence. As we shall discover, these energies are representative in the sense that each will be affected by the turbulence 
to differing degrees because the region where we place the turbulence will have a different relation to the MSW densities of these three neutrino energies. 
Since we are considering different neutrino energies it is not necessary for us to consider snapshots at other epochs from this simulation because, to first order, the profile epoch and the neutrino energy 
are degenerate: what occurs to lower energy neutrinos at early times will occur for higher energies at later times. 
For this profile we may regard the $45\;{\rm MeV}$ neutrinos as representing what we expect at the epoch when turbulence is just beginning to affect the neutrinos of a given energy, the $15\;{\rm MeV}$ as representing the effect when the H resonant channel is strongly affected, and the $5\;{\rm MeV}$ neutrinos as representing what we expect when the turbulence begins to affect both H and L resonant channels.
Fixing the neutrino energy and changing the snapshot time would be an alternative way to explore this dependence between the turbulence effects and MSW densities.
Also, the actual shape of the underlying one-dimensional profile is not very important to the turbulence effects so changing the simulation will not lead to 
qualitatively different results - the reader is referred to Lund \& Kneller \cite{2013PhRvD..88b3008L} where turbulence was put into this same $M=10.8\;M_{\odot}$ simulation from Fischer \emph{et al.} at multiple post-bounce epochs and compared with results for turbulence inserted into two other two simulations, a $M=8.8\;M_{\odot}$ and a $M=18\;M_{\odot}$, by the same authors. 
\begin{figure*}[t!]
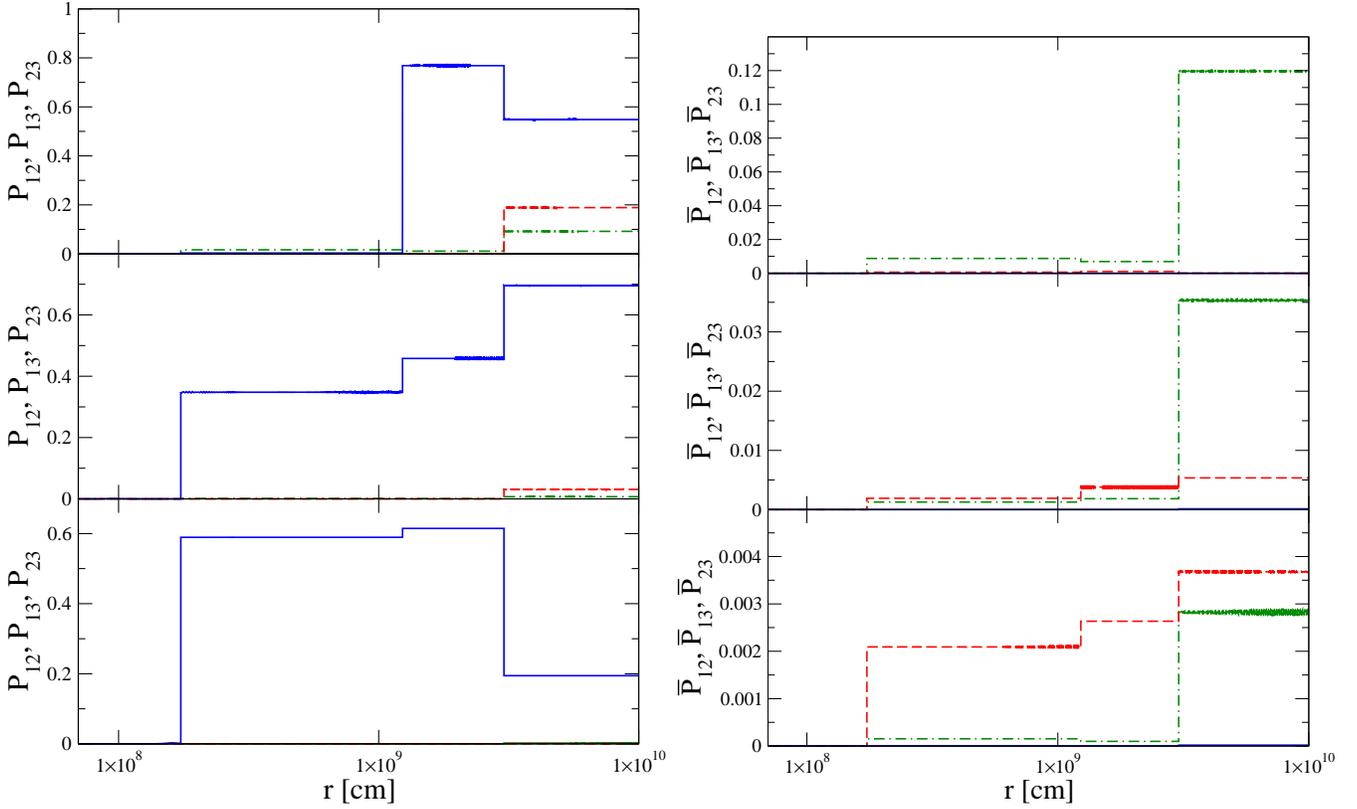
 
\includegraphics[width=0.49\linewidth]{fig2a.eps}
\includegraphics[width=0.49\linewidth]{fig2b.eps}
\caption{In the left figure we show the evolution of the neutrino transition probability $P_{12}$ (dashed dot), $P_{13}$ (dashed) and $P_{23}$ (solid) 
with energies of $5\;{\rm MeV}$ (top panel) $15\;{\rm MeV}$ (middle panel) and $45\;{\rm MeV}$ (bottom panel) through 
the profile shown in figure (\ref{fig:profile}). The right figure we show the evolution of the antineutrino transition probabilities $\bar{P}_{12}$ (dashed dot), $\bar{P}_{13}$ (dashed) and $\bar{P}_{23}$ (solid) 
with energies of $5\;{\rm MeV}$ (top panel) $15\;{\rm MeV}$ (middle panel) and $45\;{\rm MeV}$ (bottom panel) through the same profile.)\label{fig:Pevolution}}
\end{figure*}
Returning to figure (\ref{fig:profile}), the reader will observe there are three discontinuities within the profile: the reverse shock at $r_r =1,734\;{\rm km}$, the contact 
discontinuity at $r_c =12,348\;{\rm km}$ and the forward shock at $r_f=30,323\;{\rm km}$. These features were steepened by hand from the original simulation 
data: see Lund \& Kneller \cite{2013PhRvD..88b3008L} for a discussion of why this steepening was necessary. 

The evolution of neutrinos and antineutrinos with energies $5\;{\rm MeV}$, $15\;{\rm MeV}$ and $45\;{\rm MeV}$ and the given set of mixing parameters through the 
base profile are shown in figure (\ref{fig:Pevolution}). For the neutrinos the mixing between $\nu_2$ and $\nu_3$ dominates and note that the sudden discontinuities in the 
transition probability $P_{23}$ occur at the discontinuities in the profile and not at the MSW resonances (unless the two coincide). For $5\;{\rm MeV}$ neutrinos there is a noticeable 
change in both $P_{12}$ and $P_{13}$ at the forward shock; for the $15\;{\rm MeV}$ the change in the same probabilities at the same location is much smaller and 
by $45\;{\rm MeV}$ the change in $P_{12}$ and $P_{13}$ are minuscule. In the antineutrinos we again see a noticeable jump in $\bar{P}_{12}$ at the forward shock of similar size to the 
change of $P_{12}$ for neutrinos of the same energy while the jump in $\bar{P}_{12}$ at higher energies is much smaller; any jump in $\bar{P}_{23}$ is very small at all antineutrino energies (the mixing in this channel is suppressed if the mixing in $\nu_2-\nu_3$ is strong); and similarly the jumps in $\bar{P}_{13}$ are small for all energies. For future reference, the values of the transition probabilities $\{P_{12},P_{13},P_{23}\}$ 
and $\{\bar{P}_{12},\bar{P}_{13},\bar{P}_{23}\}$ through this profile for $5\;{\rm MeV}$ neutrinos are $\{0.0918,0.189,0.548\}$ and $\{0.120,7.09\times 10^{-5},3.12\times 10^{-5}\}$ respectively, for $15\;{\rm MeV}$ neutrinos they are $\{7.13\times 10^{-3},0.0308,0.696\}$ and $\{0.0353, 5.36\times 10^{-3},1.35\times 10^{-4}\}$, for $45\;{\rm MeV}$ neutrinos they are $\{2.22\times 10^{-3},6.73\times 10^{-4},0.194\}$ and $\{2.81\times 10^{-3}, 3.68\times 10^{-3},1.85\times 10^{-5}\}$. The evolution of the neutrinos and antineutrinos through the underlying base profile will determine the extent to which turbulence is able to modify the emerging probabilities. 
In general we find that if $P_{ij}$ is close to the limits of zero or unity then the effect of turbulence tends to be smaller, everything else being equal. Thus we should 
expect big effects in $P_{23}$ even at small turbulence amplitudes while effects in $P_{12}$, $P_{13}$ and the antineutrinos will require somewhat larger density fluctuations. 

It is in the region between the forward and reverse shocks that strong turbulence is seen in multi-dimensional simulations \cite{2007A&A...467.1227A,2008PhRvD..77d5023K,2013arXiv1310.7488B} so that is 
where we shall modify the profile to insert the turbulence. As in Lund \& Kneller \cite{2013PhRvD..88b3008L}, we use two random field realizations: one for the zone between the forward shock and the contact discontinuity, and a 
second between the contact discontinuity and the reverse shock. Realizations are generated by multiplying $\langle n_e(r)\rangle$ by a factor $1+F(r)$ where $F(r)$ is a Gaussian random field with 
a power spectrum $E$. Quite generally we may write the random field $F(r)$ within the region $r_<$ to $r_>$ as a Fourier series of the form
\begin{equation}\label{eq:F1D}\begin{split}
F(r)&=C_{\star}\,\tanh\left(\frac{r-r_<}{\lambda_{\star}}\right)\,\tanh\left(\frac{r_>-r}{\lambda_{\star}}\right) \\
& \times\sum_{a=1}^{N_q}\,\sqrt{V_{a}}\left\{ A_{a} \cos\left(q_{a}\,r\right) + B_{a} \sin\left(q_{a}\,r\right) \right\}. 
\end{split}
\end{equation}
The purpose of the two $\tanh$ functions is to 
damp the fluctuations close to $r_<$ and $r_>$ and the parameter $\lambda_{\star}$ is the damping scale which we set to $\lambda_{\star}=100\;{\rm km}$. 
The parameter $C_{\star}$ is the root-mean-square amplitude of the field and we shall use the same value of $C_{\star}$ for the two realizations for simplicity. 
The set of co-efficients $\{A\}$ and $\{B\}$ are independent standard, zero-mean Gaussian random variates (which ensures the mean value of $F$ is zero), 
the wavenumbers for the Fourier modes are $q_{a}$ and the quantities $V_{a}$ are `volume' co-efficients. 
The method for generating a realization of the turbulence is the same `variant C' of the Randomization Method 
described in Kramer, Kurbanmuradov, \& Sabelfeld \cite{2007JCoPh.226..897K} and used in Kneller \& Mauney \cite{2013PhRvD..88b5004K}.
The space of wavenumbers is divided into $N_{q}$ regions and from each region we select a random wavenumber $q$ using the normalized power-spectrum, $E(q)$, as a
probability distribution. The volume parameters $V_{a}$ are the integrals of the power spectrum over each region. In order to produce random fields that affect the neutrinos 
we must cover a sufficiently large dynamic range of scales. Given the size of the turbulence regions shown in figure (\ref{fig:profile}) and 
the neutrino oscillation lengthscale of $\sim 10\;{\rm km}$ at these densities, the dynamic range is found to be of order 40-50 decibels which requires at a minimum that $N_{q}$ also 
be in the range 40-50 \cite{2013PhRvD..88d5020K}. 

The power spectrum of the random field is taken to be an inverse power law of the form 
\begin{equation}
E(q) = \frac{(\alpha-1)}{2\,q_{\star}} \left( \frac{q_{\star}}{|q|}\right)^{\alpha}\, \Theta(|q|-q_{\star}). \label{eq:E1D}
\end{equation}
Here $\alpha$ is the spectral index and $q_{\star}$ is the long wavelength, small wavenumber, cutoff. 
The parameter $q_{\star}$ is sometimes called the `driving scale' since it is the longest, non-zero turbulence wavelength. 
In our case this wavelength is twice the size of the turbulence domains, that is, $1/q_\star = 2\,(r_> -r_<)$.
The assumption that the power spectrum $E$ has no spatial dependence is the simplest choice we can make but it's an 
assumption that should be examined further. Though there is evidence that the power spectrum of the angular kinetic energy during the accretion phase 
does show radial dependence \cite{2011ApJ...742...74M,BlondinReyes}, that does not automatically mean we should find radial dependence in the turbulence power spectrum during the cooling phase 
when the turbulence is far out in the stellar mantle. Similarly we know of no evidence that the turbulence power spectrum in a three dimensional simulation of a supernova at the appropriate epoch over the range of lengthscales we require is anything other than a single inverse power law: the broken power law found by Borriello et al \cite{2013arXiv1310.7488B} is relic of the two dimensional nature of the simulation they analysed.

With the Hamiltonian including turbulence constructed our plan is to generate multiple realizations of the turbulence and then solve equation (\ref{eq:S}) for the  evolution matrix for each realization. 
This approach will allow us to construct ensembles of results which we can then characterize with frequency distributions or with distribution moments.


\begin{figure}[t]
\includegraphics[clip,width=\linewidth]{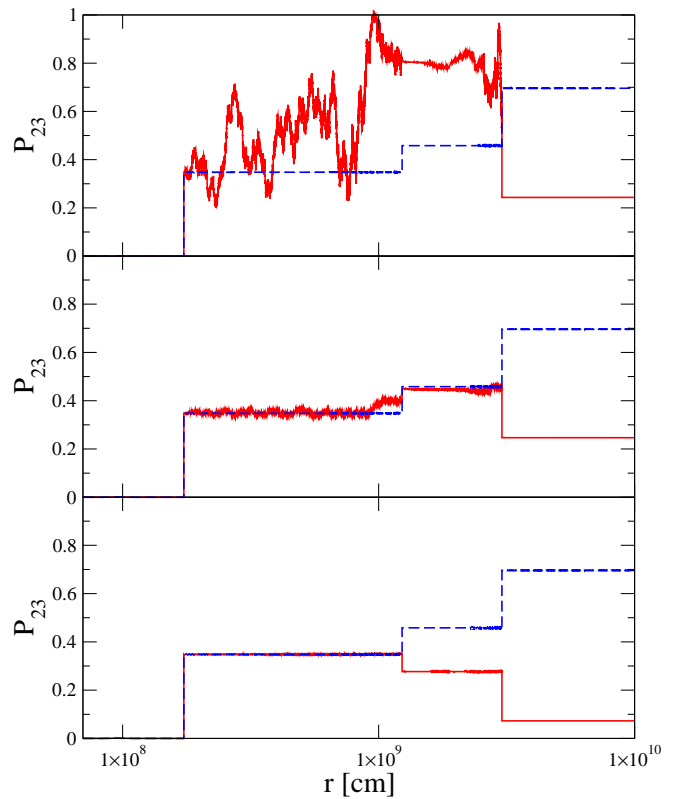}
\caption{The evolution of $P_{23}$ as a function of distance $r$ for a $15\;{\rm MeV}$ neutrino. The dashed line is the evolution 
through the underlying profile and the solid line is the evolution with a single realization of turbulence with spectral index $\alpha=5/3$. 
In the top panel $C_{\star}=10\%$, in the middle $C_{\star}=1\%$ and in the bottom $C_{\star}=0.1\%$\label{fig:P23vsr:15MeV:two effects}}
\end{figure}
\section{The two effects of turbulence}\label{sec:twoeffects}
Before presenting the results from our numerical calculations, we first take time to demonstrate the two effects of turbulence. 
As we previously stated, the first effect is the modification of the neutrino transition probability evolution 
in the region of the turbulence due to direct stimulation between the states, and the second is the modification of the transition probabilities of the neutrino as it emerges from a turbulent region if 
the transition probabilities are subject to phase effects. The two effects are neatly shown in figure (\ref{fig:P23vsr:15MeV:two effects}) where we see the evolution 
of the transition probability $P_{23}$ as a function of distance through the profile shown in figure (\ref{fig:profile}) for a $E=15\;{\rm MeV}$ neutrino. In the top panel where $C_{\star} = 0.1$ 
we see the first case: the evolution of the transition probability with turbulence differs from the evolution without turbulence as soon as the 
neutrino enters the turbulence region. In the middle panel where $C_{\star}=10^{-2}$ 
we see a small differences between the evolution with and without turbulence as soon as the neutrino enters the turbulence but the jumps at the 
discontinuities are significantly larger. In the bottom panel where $C_{\star}=10^{-3}$ 
there is no apparent difference between the evolution with and without turbulence \emph{until} the contact discontinuity. Beyond that point the two 
curves are different but exhibit no relative change until the neutrino passes through the forward shock where another large change occurs.
In these bottom two panels the change in the evolution is mainly due to a change in the phase effects; phase effects do occur in the top panel but are subordinate. But note well that that even though the turbulence 
amplitude differs by two orders of magnitude between the three calculations, the value of $P_{23}$ that emerges is very similar in each case. 

In addition to the visual distinction between the two effects, it is possible to understand these two effects in more analytic terms. 
These analytic descriptions will prove useful because they allow us to construct expectations for, and thus interpret, the results from ensembles of realizations
of the turbulence which we shall present in section \S\ref{sec:Results}.


\subsection{Phase effect distortion}

Let us first concentrate on the distortion of the phase effect. We shall use a two-flavor model and a profile with two discontinuities 
which will be found to both adequately describe our results and to understand the effect. 
In the matter basis in which we work, figure (\ref{fig:P23vsr:15MeV:two effects}) indicates that when the turbulence amplitude is small the neutrino passes through a 
set of discontinuities at the entrance and exit of the turbulent regions but evolves adiabaticaly in the turbulence region between the discontinuities. 
We can construct an evolution matrix which describes this evolution and from it derive the transition probability. 
First, the matter basis evolution matrix describing the evolution across a discontinuity located at $r$ is $S(r_+,r_-) =\tilde{U}^{\dagger}(r_+) \tilde{U}^{\dagger}(r_-)$ 
where $\tilde{U}$ is the matter mixing matrix ,$r_-$ is a point immediately before the discontinuity and $r_+$ a point immediately after. 
Adiabatic evolution of the neutrino between discontinuities means the evolution matrix must be of the form 
\begin{equation}
 S(r_b,r_a) = \left( \begin{array}{cc} \exp(-\imath \phi_1) & 0 \\ 0 & \exp(-\imath \phi_2) \end{array} \right)
\end{equation} 
where 
\begin{equation}
\phi_j=\int_{r_a}^{r_b}\,\tilde{k}_{j}(r)\,dr 
\end{equation} 
and $\tilde{k}_j(r)$ is the instantaneous eigenvalue for matter state $j$. 
Thus the evolution matrix describing the neutrino evolution through a profile with two discontinuities, located at $r_a$ and $r_b$ is  
\begin{eqnarray}
S & \sim & \tilde{U}^{\dagger}(r_{b+}) \tilde{U}^{\dagger}(r_{b-}) \left( \begin{array}{cc} \exp(-\imath \phi_1) & 0 \\ 0 & \exp(-\imath \phi_2) \end{array} \right) \nonumber \\
& & \times \tilde{U}^{\dagger}(r_{a+}) \tilde{U}^{\dagger}(r_{a-}) \label{eq:Sphaseeffect}
\end{eqnarray} 
where we have omitted the evolution up to $r_{a-}$ and beyond $r_{b+}$ assuming it to be adiabatic. 
If we denote by $P_a$ and $P_b$ the crossing probabilities through the discontinuities separately, we find 
the crossing probability for the neutrino after passing through the entire profile is 
\begin{eqnarray}
P&=&P_a\,(1-P_b)+(1-P_a)\,P_b \nonumber \\ &&+ 2\sqrt{P_a\,P_b\,(1-P_a)(1-P_b)}\,\cos(\Phi). \label{eq:phaseeffect}
\end{eqnarray} 
where $\Phi = \phi_1 - \phi_2 + constant$. Note well that $P_a$ and $P_b$ are constants and not changed when we insert turbulence into the profile and that 
$P_a \propto |\tilde{U}_{e1}^{\star}(r_{a+}) \tilde{U}_{e2}(r_{a-})|^{2}$ and $P_b \propto |\tilde{U}_{e1}^{\star}(r_{b+}) \tilde{U}_{e2}(r_{b-})|^{2}$. 
This dependence of $P$ upon the matter mixing matrix elements indicates the crossing probability $P$ is most sensitive to the turbulence via the distorted phase effect when the 
the MSW resonance density is similar to one of the densities on either side of the discontinuity. 

The place where the turbulence enters equation (\ref{eq:phaseeffect}) is via the phase difference $\Phi$. 
Every realization of the turbulence will give a different value for the phase difference and if we generate many realizations then we generate a distribution of phase differences. 
Thus we can model the effect of the turbulence by treating the phase difference $\Phi$ as a random variate drawn from a distribution $f(\Phi)$. The model we adopt for the distribution 
of $\Phi$ is the von Mises distribution which is of the form 
\begin{equation}
f(\Phi)  = \frac{\exp(\kappa \cos(\Phi-\Phi_0) )}{2\pi I_0(\kappa) }
\end{equation}
where $\Phi_0$ is the mean value of $\Phi$ and $\kappa$ is the concentration. If we define $P_{\star} = P_a\,(1-P_b)+(1-P_a)\,P_b$ and $\Delta = 2\sqrt{P_a\,P_b\,(1-P_a)(1-P_b)}$ and note that 
$|dP/d\Phi| = \Delta \sin \Phi$ then we derive the distribution for $P$ must be  
\begin{eqnarray}
f(P) & = & \frac{1}{2\pi I_0(\kappa)\sqrt{\Delta^2-(P-P_{\star})^2}} \nonumber \\ && \times \exp\left( \kappa\cos\left[\cos^{-1}\left(\frac{P-P_{\star}}{\Delta}\right)-\Phi_0\right]\right)
\end{eqnarray}
on the interval $P_{\star} -\Delta \leq P \leq P_{\star} +\Delta$.
In the limit where $\kappa \rightarrow 0$, the distribution of $\Phi$ is rectangular and the probability distribution for $P$ becomes the arcsine distribution i.e.~
\begin{equation}
f(P) = \frac{1}{2\pi\sqrt{\Delta^2-(P-P_{\star})^2}}
\end{equation}
which has a mean of $P_{\star}$ and variance of $V(P) = \Delta^2/2$. Note that in this limit there is no explicit dependence of $f(P)$ upon the spectral index $\alpha$ nor 
$C_{\star}$ because $P_{\star}$ and $\Delta$ are independent of the turbulence. 

In the other limit where the concentration is large we can expand the phase $\Phi$ around $\Phi_0$ so that to lowest order (assuming $\sin\Phi_0 \neq 0$)
\begin{equation}
P - P_0 = -\Delta \sin(\Phi_0)\, (\Phi-\Phi_0)
\end{equation} 
where $P_0 = P_\star + \Delta \cos \Phi_0$. This equation shows $P$ and $\Phi$ are linearly related in this limit and so the standard deviation of $P$ is proportional to the
the standard deviation of $\Phi$: $\sigma_P \propto \sigma_{\Phi}$. The phase difference $\Phi$ between two discontinuities is given by
\begin{equation}
\Phi \propto \int_{r_a}^{r_b}\,\left(\tilde{k}_{i}(r) - \tilde{k}_{j}(r) \right)\,dr 
\end{equation} 
and for neutrinos far from a MSW resonance - such as the $45\;{\rm MeV}$ neutrinos in the $\nu_2-\nu_3$ mixing channel - the difference between $\tilde{k}_{2}$ and $\tilde{k}_{3}$ is approximately the MSW potential $\langle V \rangle (1+F)$ in the region where we place the turbulence. This means we can write an expression for the variance of $\Phi$ which is
\begin{equation}
\langle (\Phi - \Phi_0)^2 \rangle \approx \langle \left( \int_{r_a}^{r_b}\,\langle V \rangle(r)  F(r)  \,dr \right)^2 \rangle
\end{equation} 
The integral is dominated by the longer wavelengths, lowest wavenumbers of the random field $F$ - the integral over the turbulence modes with 
wavelengths smaller than the scale height of the potential will be very small. 
If we ignore all the Fourier modes except the lowest then we predict $\langle (\Phi - \Phi_0)^2 \rangle \propto \langle A_1^2 \rangle$ and, since $q_1 \approx q_{\star}$, we have $\langle A_1^2 \rangle \propto \alpha-1$. 
Putting this together with the linear relationship between $P$ and $\Phi$ we conclude 
\begin{equation}
\sigma_{P} \propto \sqrt{ \alpha -1.}
\end{equation} 
This result indicates harder turbulence power spectra have less effect upon the transition probability than softer spectra. This makes sense because for phase distortion 
the neutrinos are more sensitive to the longest wavelength of the turbulence and those amplitudes \emph{increase} with $\alpha$. 

Obviously the distorted phase effect of turbulence requires a phase effect be present in the transition probability in the absence of turbulence. 
Glancing at figure (\ref{fig:Pevolution}) we see only $P_{23}$ has a strong phase effect with two or more semi-adiabatic transitions at the three energies we are using, plus 
$P_{12}$ and $\bar{P}_{12}$ for the $5\;{\rm MeV}$ neutrinos: in all other 
mixing channels the jumps in $P_{ij}$ at the discontinuities are too small. Thus we expect to see the distorted 
phase effect of turbulence for all three energies we are using in the $\nu_2 - \nu_3$ mixing channel, and maybe a small distorted 
phase effect in $\nu_1 - \nu_2$ \emph{and} $\bar{\nu}_1 - \bar{\nu}_2$ at $5\;{\rm MeV}$.


\subsection{Stimulated transitions}

A detailed description of the direct stimulation of transitions between neutrino states by turbulence can be found in 
Kneller, McLaughlin \& Patton \cite{Kneller:2012id} and the two papers by Patton, Kneller \& McLaughlin \cite{2014PhRvD..89g3022P,2015PhRvD..91b5001P}. 
In this description the effect of the turbulence can be understood as similar to the effect of a laser upon a polarized molecule. Comparison 
between numerical and analytical solutions reveals the description to be very powerful because it is able to predict the effect of turbulence on a case-by-case basis.
For every pair of neutrino matter states $i$ and $j$ there is an associated splitting $\delta k_{ij}$ between two eigenvalues of the Hamiltonian. 
As equation (\ref{eq:F1D}) indicates, turbulence can be described with a Fourier series and one finds that transitions between the neutrino states will be occur 
if a set of integers $\{n\}$ can be found, one for each Fourier mode, such that $\delta k_{ij} + \sum_a n_a q_a \approx 0$. 
When the condition is exact, known as a parametric resonance \cite{Ermilova,Akhmedov,1989PhLB..226..341K,1996PhRvD..54.3941B,2009PhLB..675...69K}, 
the amplitude of the transition between states $i$ and $j$ is $100\%$ no matter the amplitudes of each Fourier mode. 
Where the amplitudes of the modes enter is through the distance $\lambda$ - called the transition wavelength - over which a neutrino makes the transition from state $i$ to state $j$ or vice versa. 
This distance is inversely proportional to the coupling between the two states i.e.\ $\breve{U}_{ei}\,\breve{U}_{ej}^{\star}$, where $\breve{U}_{ei}$ are elements of the 
unperturbed matter mixing matrix, and also inversely proportional to a product of Bessel functions $J_{n_a}(z_a)$ where $z_a \propto C_a / q_a$ and the integer $n_a$ is the same integer previously identified. 
$C_a$ is the amplitude of the Fourier mode $a$. 
In order for a stimulated transition to occur one must compare $\lambda$ with the density scale height $r_{\rho}$ of the underlying profile defined to be $r_{\rho} = \rho/ (d\rho/dr)$. 
Only if $\lambda < r_{\rho}$ is a transition expected, if $\lambda > r_{\rho}$ then none will occur. 
Finally, the only difference when one considers antineutrinos is that the splitting $\delta k_{ij}$ and coupling $\breve{\bar{U}}_{ei}\,\breve{\bar{U}}_{ej}^{\star}$ are computed using a Hamiltonian 
where the MSW potential switches sign. 

At small $C_{\star}$ one typically finds the integers are $n_a = 0$ for all $a$ \emph{except} for the mode whose wavenumber is closest to the eigenvalue splitting $\delta k_{ij}$ \cite{2015PhRvD..91b5001P}. 
In this limit the smallness of $C_a$ indicates $z_a$ for that mode will also be small so we may use $J_1 \sim z$ for small $z$. 
Putting this all together we find the transition wavelength in the small amplitude limit goes as 
\begin{equation}
1/\lambda \propto C_a \breve{U}_{ei}\,\breve{U}_{ej}^{\star} / q_a.
\end{equation}
The transition wavelength $\lambda$ is inversely proportional to that resonant mode's amplitude $C_a$ - which depends upon $\alpha$ and $C_{\star}$ - and the product of 
mixing matrix elements $\breve{U}_{ei}\,\breve{U}_{ej}^{\star}$ in the region of the turbulence. In order to make $\lambda$ small and see stimulated transition we must either increase $\breve{U}_{ei}\,\breve{U}_{ej}^{\star}$ and/or $C_a$. 
Increasing $\breve{U}_{ei}\,\breve{U}_{ej}^{\star}$ means the turbulence must be in the vicinity of the MSW density because the product takes on its maximal value at that location.
This is the same requirement as in the distorted phase effect. 
Since the expectation value of $C_a$ goes as $\langle C_a\rangle^2 = C_{\star}^2 (q_{\star}/q_a)^{\alpha}$ we see increasing $C_a$ can be achieved by either increasing 
$C_{\star}$ or decreasing $\alpha$. 

The evolution of the underlying profile means the splitting between the neutrino eigenvalues $\delta k_{ij}$ and 
the coupling $\breve{U}_{ei}\,\breve{U}_{ej}^{\star}$ between the matter states both evolve with $\langle V\rangle$. This means neutrinos experience a parametric resonance only at a point and at 
present it is only possible to predict where and with what approximate strength the transitions occur. 
Phenomenologically one finds the distributions of the transition probabilities are exponentially distributed with greater 
widths as $C_{\star}$ and $\breve{U}_{ei}\,\breve{U}_{ej}^{\star}$ increase \cite{2010PhRvD..82l3004K}. 

In the limit of large turbulence amplitudes and optimally placed turbulence, stimulated transitions occur essentially continuously in the turbulence region - the parametric resonance condition is fulfilled many times through the profile - and in virtually all realizations. When this occurs one finds the evolution matrix for the evolution between the discontinuities becomes essentially random rendering the evolution across the discontinuities 
unimportant. In this strong stimulated transition limit an ensemble of evolution matrices approaches that of 
an ensemble of $N$-flavor circular unitary matrices \cite{1962JMP.....3..140D} where the distribution of every element of the matrices is identical. 
This limit is known as depolarization and from the expectation for the structure of the ensemble of  evolution matrices we can derive the distributions of the transition probabilities. First, we
note that the $N$ real components, $x_{ij}$, plus the $N$ imaginary components, $y_{ij}$, of the elements of a row or column in every  evolution matrix 
form a 2$N$ Euclidean space. The requirement of unitarity of the evolution matrix is equivalent to the definition that a vector made from these real and imaginary components lies upon the surface of 
a unit sphere in this space. Since these 2$N$ quantities are identically distributed, the probability $f$ of a particular set of the elements from a row or column must be uniform 
over the surface of the sphere. For example, if we chose to look at a column $j$ then the probability that we are located at $\{x_{1j},y_{1j},x_{2j},y_{2j},\ldots\}$, 
must be proportional to the area element $dA$ allowing us to write  
\begin{eqnarray}
&&f(x_{1j},y_{1j},x_{2j},y_{2j},\ldots)d^{N}xd^{N}y \propto dA \nonumber \\
&&\;= \delta\left(1-\sum_{i=1}^{N}x_{ij}^{2}-\sum_{i=1}^{N}y_{ij}^{2}\right)\,\prod_{i=1}^{N} dx_{ij}\,\prod_{i=1}^{N} dy_{ij}.\;\;\;\;\label{eq:dAxy}
\end{eqnarray}
If we now change variables so that each of the $N$ independent pairs $x_{ij},y_{ij}$ are expressed as
\begin{eqnarray}
x_{1j}=\sqrt{P_{1j}}\,\cos\theta_{1j}, &\;\;\;& y_{1j}=\sqrt{P_{1j}}\,\sin\theta_{1j},\\
x_{2j}=\sqrt{P_{2j}}\,\cos\theta_{2j}, &\;\;\;& y_{2j}=\sqrt{P_{2j}}\,\sin\theta_{2j},
\end{eqnarray}
then the $P_{ij}$'s are found to be distributed as 
\begin{equation}
f(P_{1j},\ldots P_{N j})d^{N}P \propto \delta\left(1-\sum_{i=1}^{N}P_{ij}\right)\,\prod_{i=1}^{N} dP_{ij}.\label{eq:dA}
\end{equation}
The set of transition probabilities $\{P_{1j},\ldots P_{N j}\}$ are uniformly distributed on the surface of a standard $N-1$ simplex. 
Equation (\ref{eq:dA}) can be integrated over $N-1$ of the $P_{ij}$'s and normalized so that we derive the final result that element $P_{ij}$ must
be distributed according to 
\begin{equation}
f(P_{ij}) =(N-1)\,(1-P_{ij})^{N-2}. 
\end{equation} 
The shape of the distribution is controlled by the number of flavors $N$ that are involved and nothing else.
With the distribution for $P_{ij}$ found it is a simple task to determine that the mean and variance are 
\begin{eqnarray}
\langle P_{ij}\rangle &=&\frac{1}{N},\\
V(P_{ij})&=&=\sigma_{ij}^{2} = \frac{N-1}{N^2\,(N+1)}
\end{eqnarray} 
For the specific case of $N=2$ the distribution is uniform with mean $1/2$ and variance $1/12$: for $N=3$ the distribution is triangular with mean $1/3$ and variance $1/18$. 
Note that the depolarized limits do not explicitly depend upon any property of the turbulence, they are functions only of the appropriate number of flavors involved. It is 
via $N$ that the turbulence amplitude and spectral index will enter because the appropriate value of $N$ will change as $C_{\star}$ and $\alpha$ are varied.


Since there are three flavors of neutrino it would seem we should use the $N=3$ case but in practice whether 3-flavor depolarization is actually reached depends upon the placement of the turbulence in the profile in relation 
to the H \emph{and} L resonance densities for a given energy. If not located in the appropriate place in the profile for a given neutrino energy, two flavor depolarization may be more appropriate.
As we stated, the distance over which a neutrino makes the transition from matter state $j$ to matter state $i$ is proportional to the 
product of instantaneous mixing matrix elements $\breve{U}_{ei}\breve{U}_{ej}^{\star}$. 
This product has its maximal value at the resonance between states $i$ and $j$. Figure (\ref{fig:profile}) shows that the H resonance density for 
the $E=45\;{\rm MeV}$ neutrinos is below the densities where we place the turbulence by a factor $\gtrsim 3$, the L resonance density is lower by a factor $\gtrsim 300$. 
For $E=45\;{\rm MeV}$ neutrinos we should expect some difficultly stimulating transitions between matter states $\nu_1$ and $\nu_2$ but it should be somewhat easier for 
mixing between states $\nu_2$ and $\nu_3$. For the lower energy of $15\;{\rm MeV}$ the H resonance density is very similar to the density of the profile between 
the reverse shock and the contact discontinuity which would lead us to expect a strong stimulated transition effect in this channel for this energy. 
The L resonance for this same energy again lies below the density of the profile meaning the product of mixing matrix elements $\breve{U}_{e1}\breve{U}_{e2}^{\star}$ will be small 
again suppressing stimulated transitions between $\nu_{1}$ and $\nu_{2}$. Finally, the H resonance density for the $5\;{\rm MeV}$ neutrinos is similar to the 
density of the profile between the contact discontinuity and the reverse shock \emph{and} the difference between the L resonance density and the profile 
density between the reverse shock and the contact discontinuity is only of order a factor of a few. 
Thus of the three energies we are considering, the $5\;{\rm MeV}$ neutrinos have the best prospect of exhibiting stimulated transitions 
between all three states and reaching 3 flavor depolarization, the $E=15\;{\rm MeV}$ and $E=45\;{\rm MeV}$ neutrinos should show evidence for stimulated transition only between 
two states. 


\begin{figure*}[t!]
\includegraphics[clip,width=\linewidth]{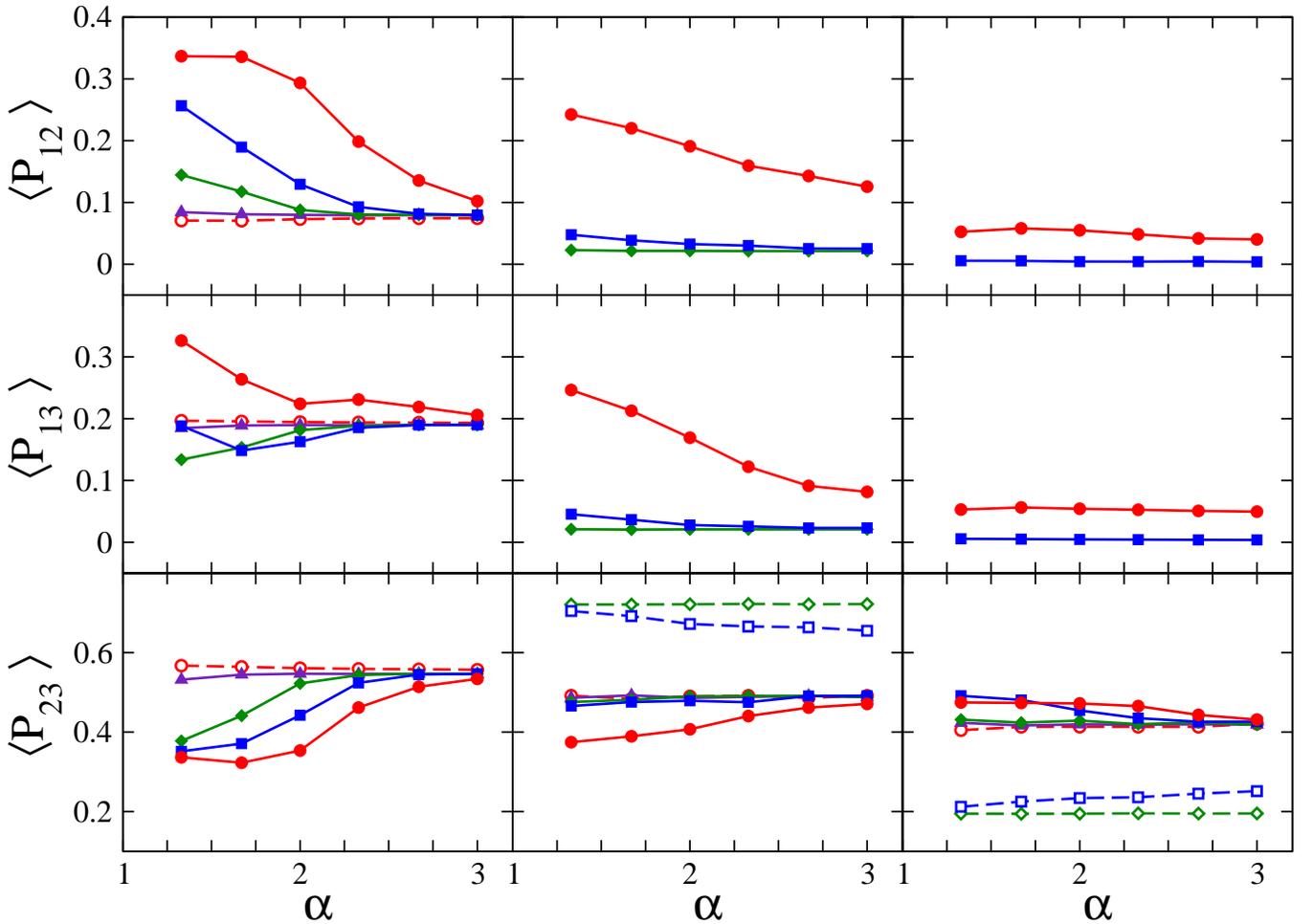}
\caption{The mean values of $P_{12}$ (top row), $P_{13}$ (middle row) and $P_{23}$ (bottom row) as a function of the power spectral index.The leftmost column is for $5\;{\rm MeV}$ neutrinos, 
the middle for $15\;{\rm MeV}$ and the rightmost is  $45\;{\rm MeV}$. In all panels the curves correspond to $C_{\star}=0.5$ (filled circles), $C_{\star}=0.3$ (filled squares), 
$C_{\star}=0.1$ (filled diamonds), $C_{\star}=10^{-2}$ (filled triangles),  $C_{\star}=10^{-3}$ (open circles), $C_{\star}=10^{-4}$ (open squares) and  $C_{\star}=10^{-5}$ (open diamonds).
For the sake of clarity, not all lines appear in each panel; where a line is missing it should be taken to be negligibly different from the smallest value of $C_{\star}$ shown in the panel.\label{fig:mean}}
\end{figure*}
\begin{figure*}[t]
\includegraphics[clip,width=\linewidth]{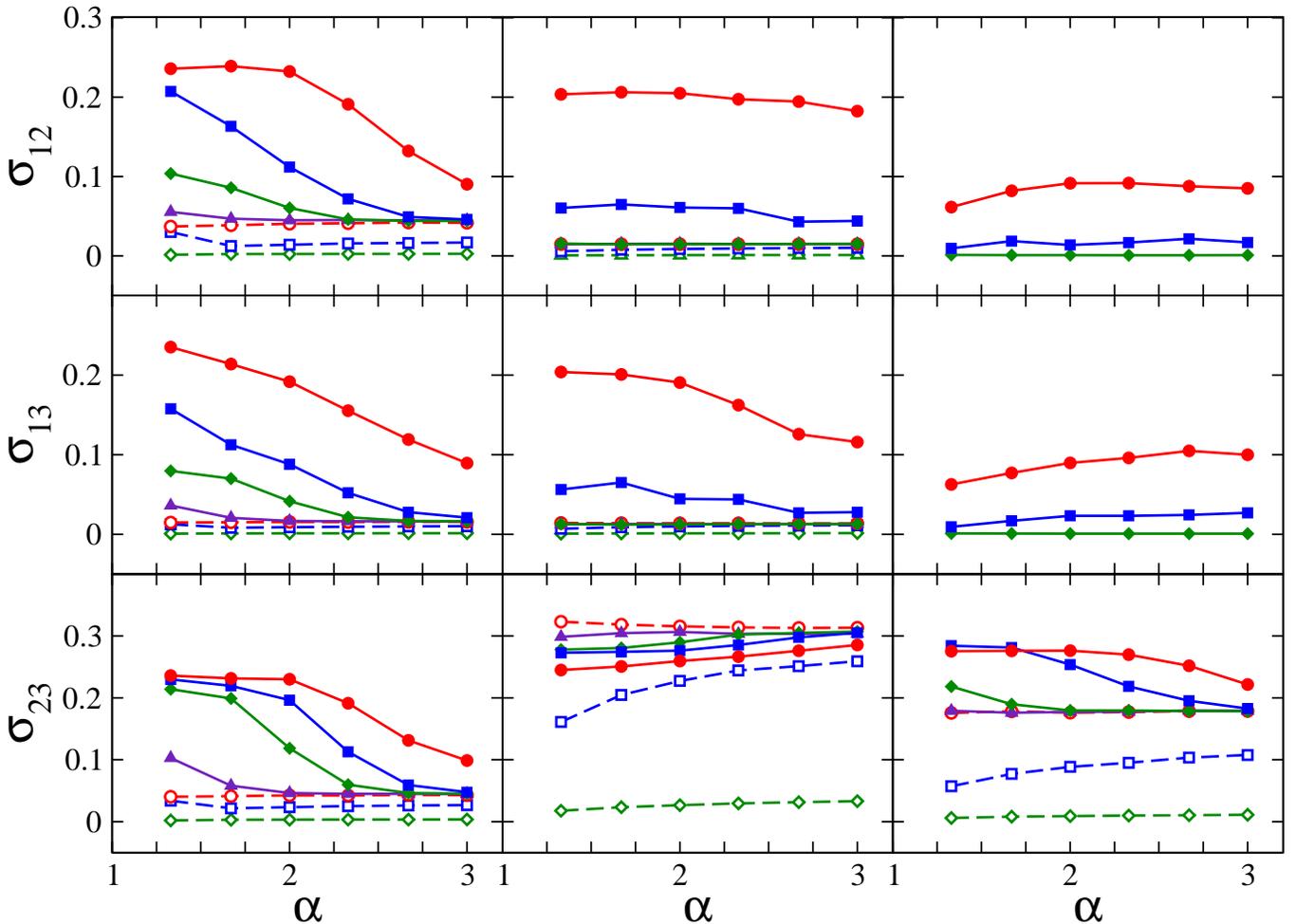}
\caption{The standard deviation of $P_{12}$ (top row), $P_{13}$ (middle row) and $P_{23}$ (bottom row) as a function of the power spectral index.The leftmost column is for $5\;{\rm MeV}$ neutrinos, 
the middle for $15\;{\rm MeV}$ and the rightmost is  $45\;{\rm MeV}$. In all panels the curves correspond to $C_{\star}=0.5$ (filled circles), $C_{\star}=0.3$ (filled squares), 
$C_{\star}=0.1$ (filled diamonds),  $C_{\star}=10^{-2}$ (filled triangles),  $C_{\star}=10^{-3}$ (open circles), $C_{\star}=10^{-4}$ (open squares) and  $C_{\star}=10^{-5}$ (open diamonds).
Again, for the sake of clarity, not all lines appear in each panel; where a line is missing it should be taken to be negligibly different from the smallest value of $C_{\star}$ shown in the panel.\label{fig:sigma}}
\end{figure*}
\begin{figure}[t]
\includegraphics[clip,width=\linewidth]{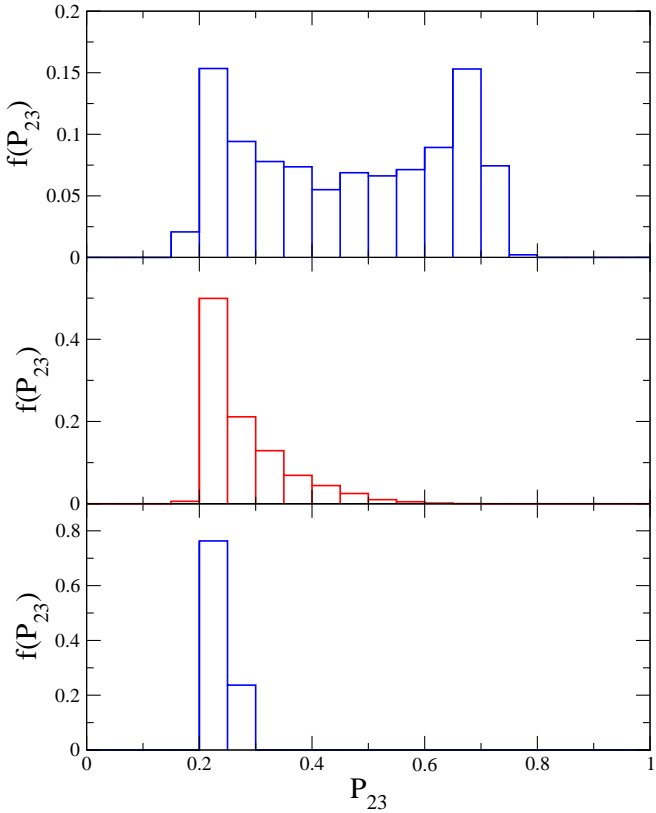}
\caption{Frequency distribution of the transition probability $P_{23}$ for the case of a $45\;{\rm MeV}$ neutrino. The spectral index is set to $\alpha = 5/3$ 
and, from top to bottom, the turbulence amplitude is $C_{\star}=10^{-3}$, $C_{\star}=10^{-4}$ and $C_{\star}=10^{-5}$. \label{fP23:45MeV.small amplitude} }
\end{figure}

The stimulated transitions work between the discontinuities in the profile while the evolution across the discontinuities themselves is still given by $\tilde{U}^{\dagger}(r_+) \tilde{U}^{\dagger}(r_-)$.
Thus stimulated transitions and distorted phases do not work separately, they work in tandem. The essential difference between the two is for stimulated transitions 
one replaces the central evolution matrix in equation (\ref{eq:Sphaseeffect}) which describes adiabatic evolution between the discontinuities with a matrix that may or may not have 
non-zero off diagonal elements depending upon whether there were any stimulated transitions in the 
turbulence region. Even if there are no stimulated transitions in the turbulence region, at a minimum the turbulence will distort the phase. 
For this reason we often find circumstances where the distributions for the transition probabilities in a given channel are not completely described by a distorted phase model nor solely described by 
stimulated transitions but rather exhibit contributions from both. These situations arise because stimulated transitions are often found to be `all-or-nothing'.  
Thus the frequency distributions of the transition probabilities in these circumstances are seen to be mixtures of two distinct, more fundamental, distributions -
see for example figure (12) in Kneller \& Volpe \cite{2010PhRvD..82l3004K} - which can cause difficulty with interpretation of results. 
Mixing between distributions will shift the means and variances of the total distribution from the expected values of the two, more fundamental, components. 
If the distribution Type A, with mean transition probability $\langle P \rangle_A$ and variance $\sigma^2_A$, contributes a fraction $f$ 
to the total distribution, and a different distribution, Type B, with mean transition probability $\langle P \rangle_B$ and variance $\sigma^2_B$ contributes $1-f$, then the mean transition 
probability of the total distribution is $\langle P\rangle = f\,\langle P \rangle_A + (1-f)\,\langle P \rangle_B$ and similarly the variance is also 
$\sigma^2 = f\,\sigma^2_A + (1-f)\,\sigma^2_B$. One should expect the fraction $f$ to depend upon the neutrino energy and snapshot time as well as the turbulence 
amplitude $C_{\star}$ and the spectral index $\alpha$. 
Knowing $f$ is useful if one wants to, say, simulate supernova neutrino signals using Monte Carlo methods. 
The neutrinos we receive at a given instant from the next supernova in our galaxy will all have traveled through similar 
turbulence as they traversed the mantle of the star which has the consequence that the transition probabilities of the neutrinos of a given energy will be very strongly correlated - the 
size of the proto-neutron in not sufficient to wash this correlation out unless the turbulence is very anisotropic \cite{2013PhRvD..88d5020K}. This means that the neutrino signal 
will not have sampled the ensemble of realizations of the turbulence, we receive neutrinos which have been affected by essentially just one realization. 
The probability this realization gives transition probabilities distributed according to distribution Type A is the fraction $f$; the probability 
the transition probabilities are distributed according to distribution Type B is $1-f$.


\section{Results}\label{sec:Results}
\subsection{Neutrinos}
In figures (\ref{fig:mean}) and (\ref{fig:sigma}) we show the mean values and standard deviation of the distributions of the 
neutrino transition probabilities $P_{12}$, $P_{13}$ and $P_{23}$  as a function of the spectral index and the turbulence amplitude for 
the three representative energies we are using. 
Even a cursory glance indicates there is a great deal of rich behavior as a function 
of the three parameters we have varied to generate the figures. Before diving into the results in depth to try and understand 
why we see the trends we do, let us summarize them:
\begin{itemize}
\item {\bf $E=45\;{\rm MeV}$}. When $C_{\star} = 50\%$, $\alpha \leq 7/3$ and $C_{\star} = 30\%$, $\alpha \leq 5/3$, the $\nu_2$-$\nu_3$ mixing channel reaches $\langle P_{23} \rangle = 0.5$ and $\sigma_{23} = 0.28$. 
At these same large amplitudes but larger $\alpha$ we see the mean and standard deviation, $\langle P_{23} \rangle$ and $\sigma_{23}$ respectively, both fall with increasing $\alpha$. 
In the range $0.1\% \lesssim C_{\star} \lesssim 10\%$ it appears $\langle P_{23} \rangle$ is fixed at $\langle P_{23} \rangle =0.44$ independent of $\alpha$ and $C_{\star}$ and similarly the standard deviation $\sigma_{23}$ is also independent of $\alpha$ and $C_{\star}$ fixed at $\sigma_{23} \approx 0.18$. These same two values appear to be the asymptotic limits of $\langle P_{23} \rangle$ and $\sigma_{23}$ for $C_{\star} = 50\%$ and $C_{\star} = 30\%$ for large $\alpha$. 
At very small turbulence amplitudes, $C_{\star} \lesssim 0.01\%$, a dependence of $\langle P_{23} \rangle$ and $\sigma_{23}$ upon $C_{\star}$ and $\alpha$ re-emerges and 
it appears both $\langle P_{23} \rangle$ and the standard deviation $\sigma_{23}$ \emph{increase} as the power spectrum becomes softer.  
In $\langle P_{12} \rangle$, $\langle P_{13} \rangle$, $\sigma_{12}$ and $\sigma_{13}$ we only see a difference from turbulence free results when $C_{\star} = 50\%$ and $C_{\star} = 30\%$
with little dependence upon $\alpha$.
\item {\bf $E=15\;{\rm MeV}$}.
For this energy the mean value of $P_{23}$ is approximately $\langle P_{23} \rangle = 0.5$ at $\alpha =3$ for all turbulence amplitudes above $C_{\star} = 0.1\%$. 
Except for the case of $C_{\star}=0.5$, there is no apparent running of $\langle P_{23} \rangle$ with $\alpha$ or $C_{\star}$ in this range of amplitudes: in $\sigma_{23}$ the trend appears 
to be a small increase in $\sigma_{23}$ with $\alpha$ from $\sigma_{23}=0.28$ at $\alpha=4/3$ to $\sigma_{23}=0.3$ at $\alpha=3$ and we observe that at fixed $\alpha$, as $C_{\star}$ decreases the standard deviation increases.  
For $C_{\star}=0.5$ the increase of $\langle P_{23} \rangle$ and $\sigma_{23}$ with increasing $\alpha$ is more obvious and we note that $\langle P_{12} \rangle$, $\langle P_{13} \rangle$, $\sigma_{12}$ and $\sigma_{13}$
are noticeably different at $C_{\star}=0.5$ from even the case of $C_{\star}=0.3$.
For $C_{\star} \lesssim 0.1\%$ the evolution of the standard deviation $\sigma_{23}$ with $C_{\star}$ reverses and now $\sigma_{23}$ decreases with $C_{\star}$. However, as with the 
$E=45\;{\rm MeV}$ neutrinos, the trend appears to be at very small turbulence amplitudes there is an increase of $\sigma_{23}$ and larger 
difference between the turbulence -free result of $\langle P_{23} \rangle$ with increasing $\alpha$. 
\item {\bf $E=5\;{\rm MeV}$}. 
For this energy we observe much larger turbulent effects in the $\nu_1$-$\nu_2$ and $\nu_1$-$\nu_3$ mixing channels than for the other two energies considered particularly 
for small $\alpha$. When $C_{\star}=50\%$ and $\alpha \leq 5/3$  the mean values of $P_{12}$, $P_{13}$ and $P_{23}$ all plateau at $\langle P_{ij} \rangle = 0.33$ 
and the standard deviations $\sigma_{12}$, $\sigma_{13}$ and $\sigma_{23}$ all reach $\sigma_{ij} = 0.23$ in the same range of amplitudes and spectral indices. 
For all amplitudes greater than $C_{\star} \gtrsim 1\%$ the trend of $\langle P_{12} \rangle$, $\langle P_{13} \rangle$ and $\langle P_{23}\rangle$ with $\alpha$ is towards a fixed value 
with simultaneous decreasing standard deviations $\sigma_{12}$, $\sigma_{13}$ and $\sigma_{23}$. 
At small amplitudes for the turbulence, $C_{\star} \lesssim 1\%$, all three transition probabilities of the $5\;{\rm MeV}$ neutrinos approach the previously reported 
values through the underlying profile. The convergence is more rapid for softer power spectra: e.g.~ at $C_{\star}=10\%$ the mean of $P_{12}$ and $P_{23}$ are measurably different 
from the turbulence free values at $\alpha=5/3$ but not so at $\alpha = 7/3$. 
\end{itemize}
Let us now examine these results in more detail and try to find explanations of the trends we observe using the two different descriptions 
for the effects of turbulence from section \S\ref{sec:twoeffects}.


\subsubsection{$E=45\;{\rm MeV}$}
We first examine the $E=45\;{\rm MeV}$ neutrinos and focus upon $P_{23}$. At large amplitudes and hard spectral indices the $\nu_2$-$\nu_3$ mixing channel appears to reach the two-flavor 
depolarized limit since $\langle P_{23} \rangle = 0.5$ and $\sigma_{23} = 0.28$ when $C_{\star} = 50\%$ and $\alpha \leq 7/3$ and 
$C_{\star} = 30\%$ and $\alpha \leq 5/3$. The decrease of $\langle P_{23} \rangle$ and $\sigma_{23}$ for the same large amplitudes but larger $\alpha$ is again in line with what we 
expect from stimulated transitions. The reduced amplitude of the modes that have wavelengths of order the eigenvalue splittings mean the stimulated transitions are not as strong and the 
distribution of $P_{23}$ will no longer be uniform. 

At smaller turbulence amplitudes our explanation for the results changes. In the range $0.1\% \lesssim C_{\star} \lesssim 10\%$ the 
mean value of $P_{23}$ and the standard deviation $\sigma_{23}$ are independent of $\alpha$ fixed at apparently arbitrary values. 
Only if we push further to even smaller turbulence amplitudes, $C_{\star} \lesssim 0.01\%$, do we see the dependence upon $C_{\star}$ and $\alpha$ re-emerge but 
when it re-emerges the trend that both $\langle P_{23} \rangle$ and the standard deviation $\sigma_{23}$ \emph{increase} as the power spectrum becomes softer.
This behavior of $\langle P_{23} \rangle$ and $\sigma_{23}$ for $C_{\star} \leq 10\%$ must be explained by using the distorted phase model so 
let us use this model to try and predict the actual values found in (\ref{fig:mean}) and (\ref{fig:sigma}) for the $E=45\;{\rm MeV}$ neutrinos in this amplitude range. 
From analyzing the evolution without turbulence - figure (\ref{fig:Pevolution}) - we find the the transition probabilities for these neutrinos at 
the reverse and forward shocks give $P_a =0.57$ and at the second $P_b =0.93$. These can be combined to give $P_{\star} = 0.44$ and $\Delta =0.25$. Since the transition probability in the absence of turbulence is 
$P_{23} = 0.20$ we deduce $\Phi = 163^{\circ}$. With this information in hand we predict the distribution of the transition probability $P_{23}$ when we insert the turbulence 
will lie in the range of $P_{\star} - \Delta = 0.19 $ to $P_{\star} + \Delta = 0.69$. When the concentration $\kappa$ is small we expect an arcsine distribution for $P_{23}$ with 
a mean $\langle P_{23} \rangle = P_{\star} = 0.44$ and standard deviation $\Delta / \sqrt{2} = 0.18$. 
These predictions match the data well so we interpret our results as meaning that in the range $0.1\% \lesssim C_{\star} \lesssim 10\%$ the $45\;{\rm MeV}$ neutrinos are experiencing a strong distorted phase effect. 
At smaller turbulence amplitudes when the concentration $\kappa$ is larger the distribution will be like a half-Gaussian because the turbulence-free value of $P_{23} = 0.20$ is close to 
the lower limit of the distribution. 

The frequency distribution of $P_{23}$ for $45\;{\rm MeV}$ neutrinos at $C_{\star}=10^{-3}$, $C_{\star}=10^{-4}$ and $C_{\star}=10^{-5}$ is shown in 
figure (\ref{fP23:45MeV.small amplitude}) and these expected shapes of the distributions is seen in the numerical results. At $C_{\star}=10^{-3}$ the 
distribution is symmetric around $P_{23}= 0.44$ peaking at the extremes $P_{23}=0.19$ and $P_{23}=0.69$ as an arcsine distribution should. 
At  $C_{\star}=10^{-4}$ and $C_{\star}=10^{-5}$ the symmetry is lost and the distribution looks more like an exponential or half-Gaussian. 
The running of $\langle P_{23} \rangle$ and $\sigma_{23}$ with $\alpha$ for the $E=45\;{\rm MeV}$ neutrinos and smaller turbulence amplitudes is also in line 
with our expectations from the distorted phase model because we see $\sigma_{23}$ increase with $\alpha$ e.g.~ $C_{\star} = 10^{-4}$. 

Let us now consider the other mixing channels at this same energy. 
Compared to $P_{23}$, the mean of the transition probabilities $\langle P_{12} \rangle$ and $\langle P_{13} \rangle$ for the $E=45\;{\rm MeV}$ neutrinos appear quite 
unremarkable differing from the turbulence free limit only when $C_{\star} = 0.5$ and then possessing only a soft dependence upon $\alpha$. The standard deviations $\sigma_{12}$ and $\sigma_{13}$ 
evolve similarly. At this energy the distorted phase effect of turbulence does not operate in these channels because the jumps in $P_{12}$ and $P_{13}$ across 
the discontinuities are small. The sensitivity to the turbulence is entirely through the stimulated transition mechanism. 
But as previously mentioned, the large separation between the turbulence densities and the L resonance MSW density at this energy means 
not every realization will cause a stimulated transition to occur in these channels so it is quite natural to expect the ensemble to be divided into two subsets. 
When we look we find this is exactly the case. The frequency distribution of the $P_{12}$ and $P_{13}$ transition probabilities 
are mixtures of a very narrow distribution which peaks at zero - the neutrinos unaffected by the turbulence - 
and an exponential distribution - the subset where stimulated transitions occurred. 

\begin{figure}[t]
\includegraphics[clip,width=\linewidth]{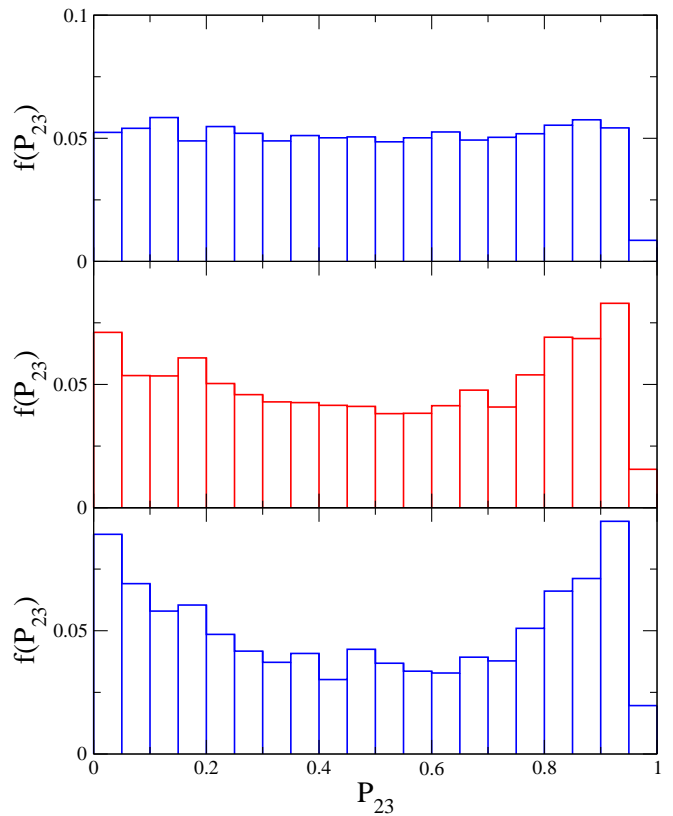}
\caption{The frequency distribution of $P_{23}$ for $15\;{\rm MeV}$ neutrinos at $\alpha= 5/3$. In the top panel $C_{\star}=10\%$, in the middle panel $C_{\star}=1\%$, 
and in the bottom panel $C_{\star}=0.1\%$, \label{fig:P23at15MeV}}
\end{figure}
\begin{figure}[t]
\includegraphics[clip,width=\linewidth]{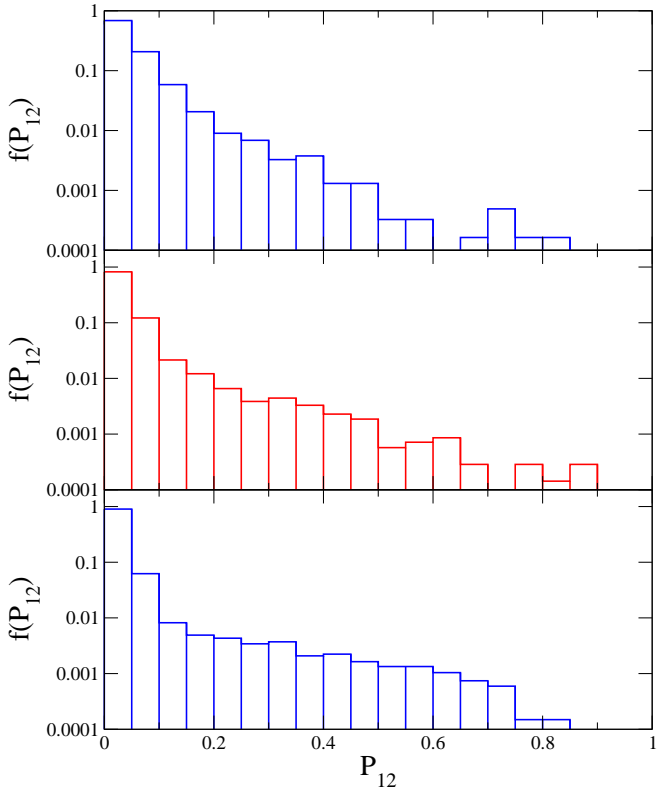}
\caption{The frequency distribution of the neutrino transition probability $P_{12}$ when $E=15\;{\rm MeV}$ and $C_{\star}=0.3$. In the top panel $\alpha=4/3$, 
in the middle $\alpha=5/3$, and in the bottom panel $\alpha =2$. \label{fig:P12:15MeV}}
\end{figure}
\begin{figure}[t]
\includegraphics[clip,width=\linewidth]{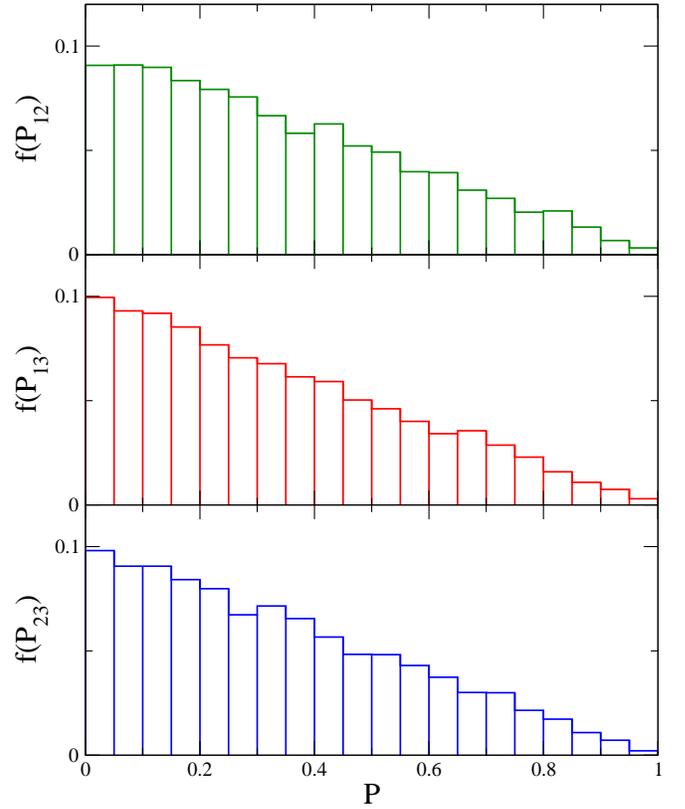}
\caption{The frequency distribution of the transition probabilities $P_{12}$, $P_{13}$ and $P_{23}$ for the case of a $5\;{\rm MeV}$ neutrino. 
The spectral index is $\alpha = 4/3$ and the turbulence amplitude is $C_{\star}=0.5$. \label{fP12,13,23:5MeV.large amplitude}}
\end{figure}
\begin{figure}[t]
\includegraphics[clip,width=\linewidth]{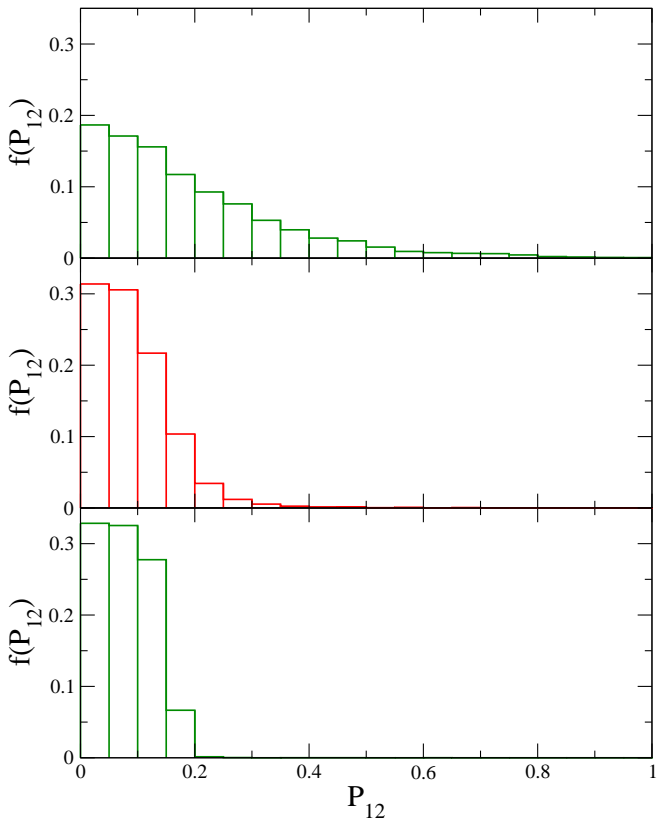}
\caption{The frequency distribution of $P_{12}$ for $E=5\;{\rm MeV}$ neutrinos at $C_{\star}=0.3$. In the top panel $\alpha=5/3$, in the middle $\alpha=7/3$, and in the bottom panel $\alpha=3$.\label{fig:fP12vsalpha}}
\end{figure}
\begin{figure*}[t]
\includegraphics[clip,width=\linewidth]{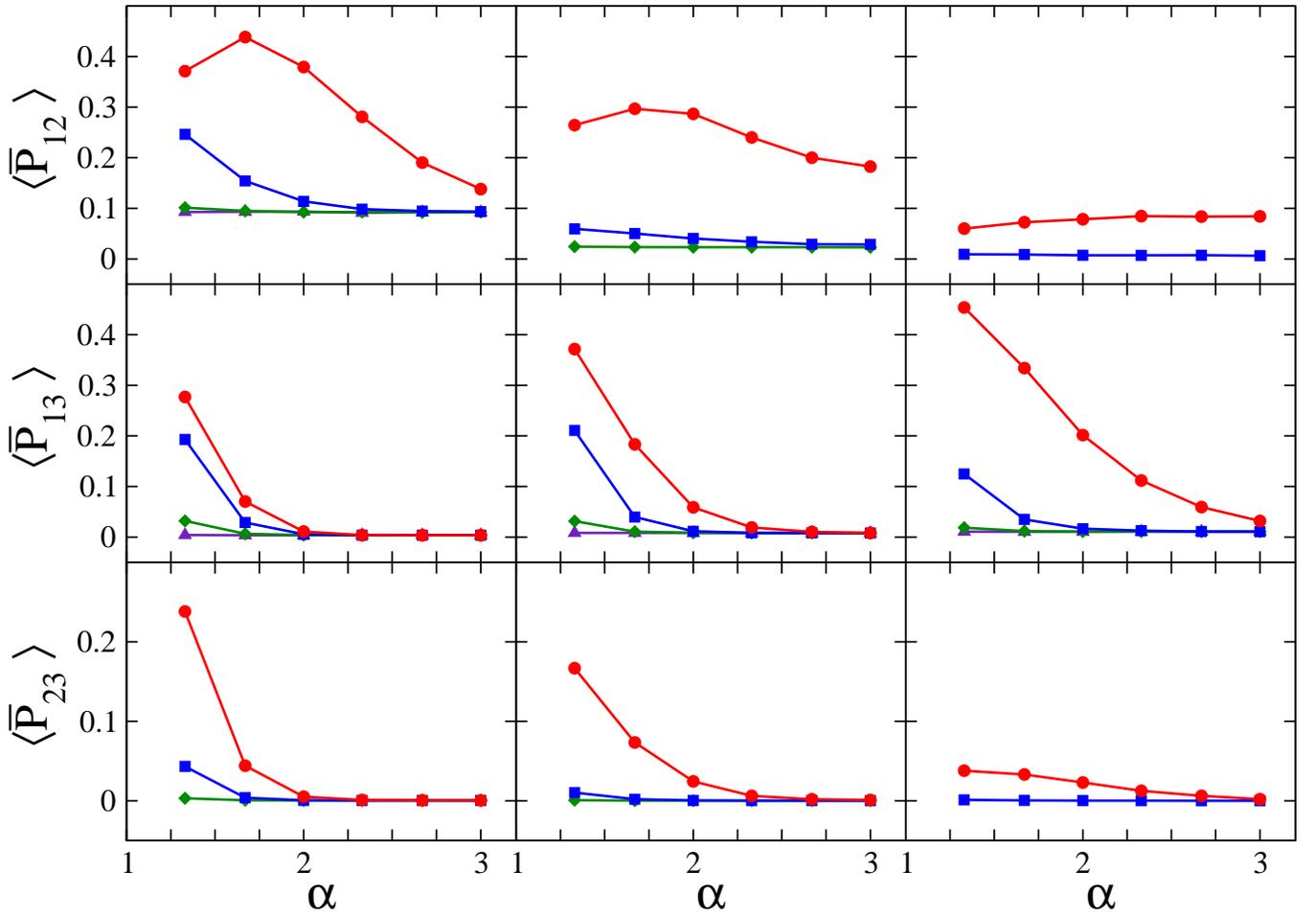}
\caption{The mean values of $\bar{P}_{12}$ (top row), $\bar{P}_{13}$ (middle row) and $\bar{P}_{23}$ (bottom row) as a function of the power spectral index.The leftmost column is for $5\;{\rm MeV}$ neutrinos, 
the middle for $15\;{\rm MeV}$ and the rightmost is  $45\;{\rm MeV}$. In all panels the curves correspond to $C_{\star}=0.5$ (filled circles), $C_{\star}=0.3$ (filled squares), 
$C_{\star}=0.1$ (filled diamonds), $C_{\star}=10^{-2}$ (filled triangles). For the sake of clarity, not all lines appear in each panel. \label{fig:meanbar}}
\end{figure*}
\begin{figure*}[t]
\includegraphics[clip,width=\linewidth]{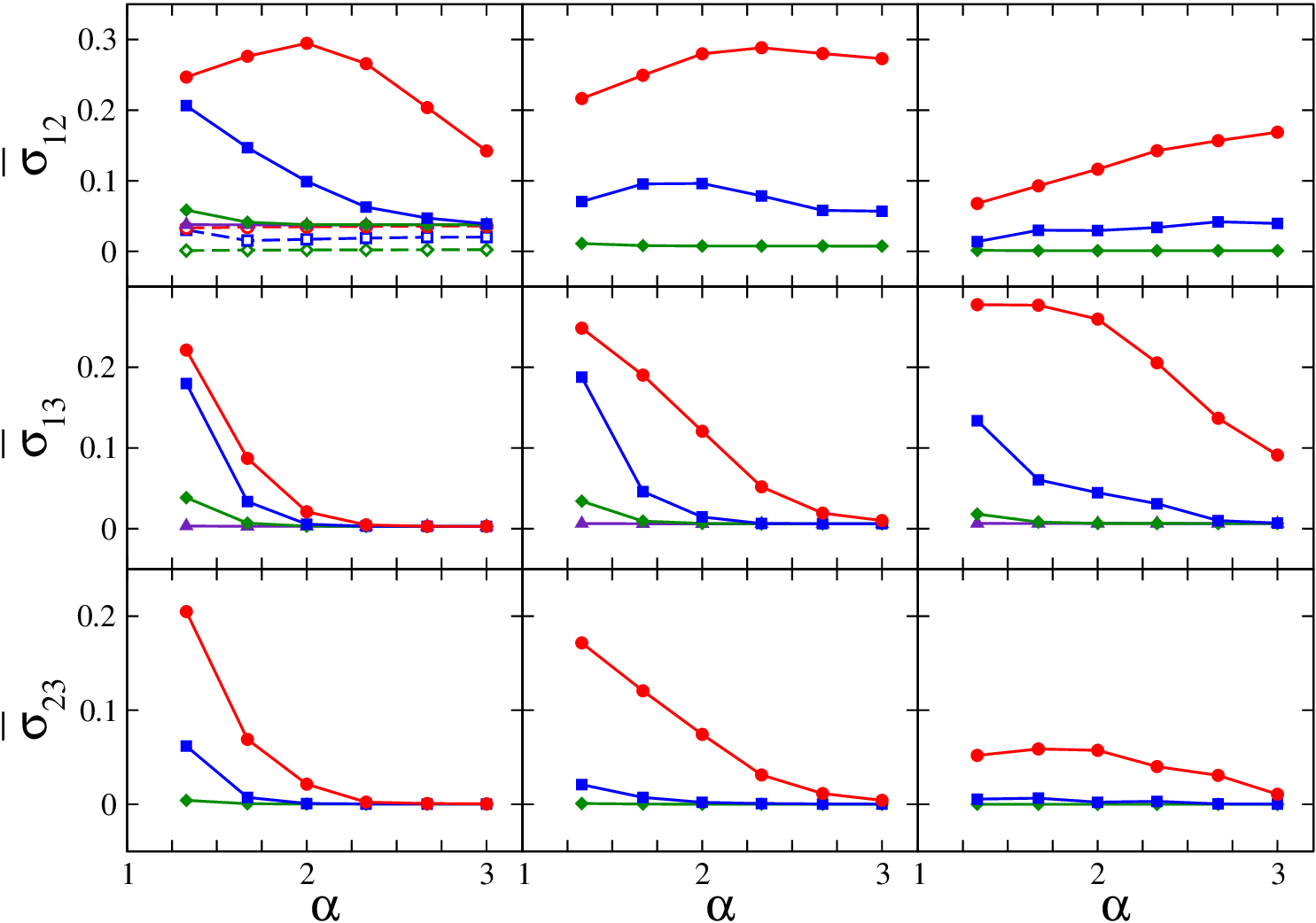}
\caption{The standard deviation of $\bar{P}_{12}$ (top row), $\bar{P}_{13}$ (middle row) and $\bar{P}_{23}$ (bottom row) as a function of the power spectral index.The leftmost column is for $5\;{\rm MeV}$ neutrinos, 
the middle for $15\;{\rm MeV}$ and the rightmost is  $45\;{\rm MeV}$. In all panels the curves correspond to $C_{\star}=0.5$ (filled circles), $C_{\star}=0.3$ (filled squares), 
$C_{\star}=0.1$ (filled diamonds),  $C_{\star}=10^{-2}$ (filled triangles),  $C_{\star}=10^{-3}$ (open circles), $C_{\star}=10^{-4}$ (open squares) and  $C_{\star}=10^{-5}$ (open diamonds).
Again, for the sake of clarity, not all lines appear in each panel. \label{fig:sigmabar}}
\end{figure*}
\begin{figure}[t]
\includegraphics[clip,width=\linewidth]{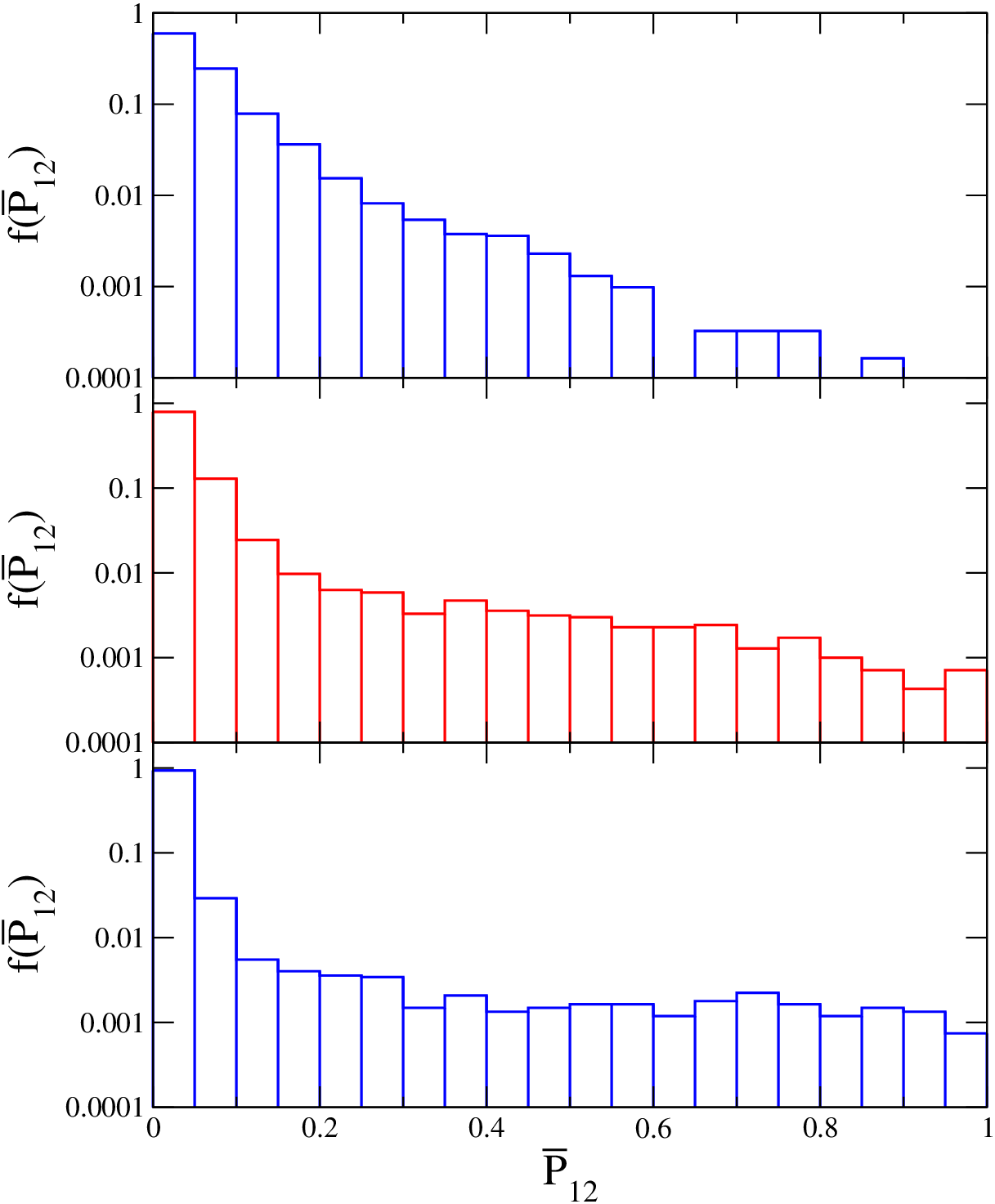}
\caption{The frequency distribution of the antineutrino transition probability $\bar{P}_{12}$ when $E=15\;{\rm MeV}$ and $C_{\star}=0.3$. In the top panel $\alpha=4/3$, 
in the middle $\alpha=5/3$, and in the bottom panel $\alpha =2$. \label{fig:P12bar:15MeV}}
\end{figure}

\subsubsection{$E=15\;{\rm MeV}$}
We now consider the energy $E=15\;{\rm MeV}$.
For this energy the evolution of the distributions of $P_{23}$ is again explained by a transition from stimulated transitions at large $C_{\star}$/small $\alpha$ to the distorted 
phase effect at smaller $C_{\star}$/ larger $\alpha$. The frequency distributions of $P_{23}$ for $15\;{\rm MeV}$ neutrinos at fixed $\alpha=5/3$ as several amplitudes $C_{\star}$ are shown 
figure (\ref{fig:P23at15MeV}). In all three cases shown the mean value of $P_{23}$ is approximately the same, around $\langle P_{23} \rangle \approx 0.5$, but the distributions 
are clearly different depending upon $C_{\star}$: for the large amplitude $C_{\star} = 0.1$ the distribution is almost uniform - 
the bin $0.95 \leq P_{23} \leq 1$ appears low - whereas the distribution for $C_{\star} = 0.001$ has a very peculiar shape with the extreme values 
of $P_{23}$ more common than the mean. These distributions reflect the two different mechanisms by which the turbulence affects the neutrinos - the 
top panel is what we would expect from stimulated transitions in the depolarized limit with two flavors whereas the distribution in the lower panels 
is the indirect, distorted phase effect since they are consistent with an arcsine distribution, similar to that seen in the 
top panel of figure (\ref{fP23:45MeV.small amplitude}), albeit on the interval of zero to unity. 

The mean values of $P_{12}$ and the standard deviations $\sigma_{12}$ shown in figures (\ref{fig:mean}) and (\ref{fig:sigma}) do not match 
the expected values from the our model distributions and this is because, as with the $E=45\;{\rm MeV}$ neutrinos, we find the ensembles 
are mixtures of distributions. 
At $C_{\star}=0.5$ inspection indicates the mixing distributions in $\nu_1-\nu_2$ for this energy are a sharp distribution which peaks at zero - the neutrinos with 
no turbulence effects - and a three flavor depolarized distribution which are the neutrinos that experienced stimulated transitions. 
But when we consider a slightly smaller value of $C_{\star}=0.3$ we find the mixing distributions for $P_{12}$ are the same narrow distribution peaked at zero but the second distribution is now 
an exponential. 
The frequency distributions of $P_{12}$ at $C_{\star}=0.3$ and $E=15\;{\rm MeV}$ at three values of $\alpha$ are shown in figure (\ref{fig:P12:15MeV}) and 
in the figure we observe a flattening of the exponentially distributed component as $\alpha$ increases.


\subsubsection{$E=5\;{\rm MeV}$}
Finally we switch to the $5\;{\rm MeV}$ neutrinos. Here neither the H resonance nor the L resonance are too far from the densities where we insert the turbulence so the
product of instantaneous mixing matrix elements $\breve{U}_{ei}\,\breve{U}_{ej}^{\star}$ are not small in any mixing channel. 
This allows stimulated transitions to occur between all three states simultaneously. 
The simultaneous mixing in all three channels indicates we might find that at sufficiently large amplitudes and hard spectral indices we could reach three-flavor depolarization. 
When we look indeed this is found for $C_{\star} \gtrsim 50\%$ and $\alpha =4/3$ shown in figure (\ref{fP12,13,23:5MeV.large amplitude}) where we see the frequency distributions of the transition probabilities 
are triangular as predicted. Thus figures (\ref{fig:mean}) and (\ref{fig:sigma}) reveal the three-flavor depolarized limit is reached for $P_{12}$ only for $C_{\star}=50\%$ and $\alpha \leq 5/3$ whereas 
the same limit appears somewhat easier to reach for $P_{23}$ because even $C_{\star} =10\%$ amplitude turbulence saturates at $\langle P_{23} \rangle =1/3$ and $\sigma_{23} = 0.23$ for $\alpha=4/3$ or at $C_{\star} =50\%$ we are able to relax the spectral index to $\alpha = 2$. This is not surprising given the location of the turbulence in the profile with respect to the $\nu_2-\nu_3$ mixing resonance density shown in figure (\ref{fig:profile}). 
Note also the figures do not indicate there are combinations of $C_{\star}$ and $\alpha$ where we achieve two flavor depolarization instead of three because we do not see $\langle P_{23}\rangle \approx 0.5 $ or $\sigma_{23} \approx 0.28$ as one would expect in that limit. 

As we move to smaller turbulence amplitudes what we observe in figures (\ref{fig:mean}) and (\ref{fig:sigma}) might seem at first to be contradictory. The mean transition probability in the 
ensembles is the same as the value in the turbulence free limit tempting one to conclude that turbulence has no effect, but the standard deviations do not support this 
conclusion because it is not until $C_{\star} \lesssim 0.001\%$ that the turbulence effect actually disappears. The contradictions can be resolved when one realizes what the 
figures are showing is that between $0.001\% \lesssim C_{\star} \lesssim 1\%$ the distributions of the transition probabilities are simply \emph{centered} on the turbulence-free limits. 
As with the $E=45\;{\rm MeV}$ neutrinos, the acute sensitivity to $C_{\star}$ is due to the distorted phase effect with the twist that now both $P_{23}$ and $P_{12}$ (and $\bar{P}_{12}$) are affected. 

The distributions of $P_{12}$ and $P_{23}$ when $0.001\% \lesssim C_{\star} \lesssim 10\%$ are mixtures of an exponential and the distorted phase distributions and 
the evolution of $\langle P_{12}\rangle$ and $\langle P_{23}\rangle$ are actually due to the evolution of the exponentially distributed subset, not the subset where the turbulence 
only distorts the phase. This can be seen in figure (\ref{fig:fP12vsalpha}) where we show the distributions for $P_{12}$ for the $E=5\;{\rm MeV}$ neutrinos 
at three values of $\alpha$ when $C_{\star}=0.3$. Note the similarity of the low end of the distribution for $\alpha=7/3$ and $\alpha=3$.


\subsection{Anti-neutrinos}

Even though we are considering just a normal hierarchy, large amplitude turbulence certainly does affect the antineutrinos.
In figures (\ref{fig:meanbar}) and (\ref{fig:sigmabar}) we show the results for the means and standard deviations of the 
ensembles of antineutrino transition probabilities. 
Let us again try to summarize what we observe in the figures. 
\begin{itemize}
 \item The antineutrino channel which is most sensitive to the turbulence is $\bar{P}_{12}$ and this sensitivity is similar to the sensitivity of $P_{12}$ seen 
in figures (\ref{fig:mean}) and (\ref{fig:sigma}). $\bar{P}_{13}$ and $\bar{P}_{23}$ are quite different from $P_{13}$ and $P_{23}$. 
 \item We also observe that at the largest value of $C_{\star}$ shown, the evolution of $\langle \bar{P}_{13}\rangle$ and $\bar{\sigma}_{13}$ with the antineutrino 
energy appears to be counter that of $\langle \bar{P}_{12}\rangle$, $\langle \bar{P}_{23}\rangle$, $\bar{\sigma}_{12}$ and $\bar{\sigma}_{23}$. 
\end{itemize}
We now try to understand these results. First, a description of the turbulence effects upon the antineutrinos for a normal hierarchy in terms of MSW resonances 
would obviously not work well because there are no resonances in the antineutrino mixing channels. Second, except for $\bar{P}_{12}$ at $E=5\;{\rm MeV}$, 
the distorted phase effect will not be prominent because the adiabaticity of the transitions for the antineutrinos across the discontinuities in the profile are large. 
It is this lack of distorted phase effects that explains the greatly reduced sensitivity of the neutrinos to the turbulence amplitude. 
The absence of distorted phase effects in the majority of the results shown in  figures (\ref{fig:meanbar}) and (\ref{fig:sigmabar}) makes their interpretation much easier 
than the neutrino transition probabilities. 
The only explanation that applies is that stimulated transitions.

If we look closely we see for the case of $\bar{P}_{12}$ when $E=5\;{\rm MeV}$ and $E=15\;{\rm MeV}$ we see that $\bar{P}_{12}$ appears to be as sensitive to the 
turbulence as the neutrinos in the $P_{12}$ channel. This can be explained by the stimulated transition model. In the turbulent region the eigenvalue splitting $\delta k_{12}$ and 
$\delta \bar{k}_{12}$ are both approximately equal to the MSW potential $V_{ee}$ and so the coupling between the states, $\breve{U}_{ei}\,\breve{U}_{ej}^{\star}$ and $\breve{\bar{U}}_{ei}\,\breve{\bar{U}}_{ej}^{\star}$
are also approximately equal. This equivalence means it should be as easy to stimulate a transition between states $\bar{\nu}_{1}$ and $\bar{\nu}_{2}$ as it is between 
${\nu}_{1}$ and ${\nu}_{2}$ so the response to the turbulence will be the same. 

But in all other cases the antineutrinos are not as sensitive to the turbulence as the neutrinos which, again, can be explained stimulated transition description. The difficulty of stimulating transitions between antineutrino states is twofold: first, in the normal hierarchy, the splitting between the eigenvalues are larger which means we require shorter wavelength Fourier modes in order 
to fulfill the parametric resonance condition and, with an inverse power law power spectrum, the amplitudes of these modes are smaller. Secondly 
the coupling between the states, $\breve{\bar{U}}_{ei}\,\breve{\bar{U}}_{ej}^{\star}$, is also generally smaller in the antineutrinos. Both smaller amplitudes for the resonance modes and smaller coupling lengthen 
the distance $\lambda$ over which the transition occurs. Since we need $\lambda$ to be smaller than the density scale height $r_{\rho}$ in order to see a stimulated transition, 
it requires very large $C_{\star}$ before stimulated transitions appear in the antineutrino mixing channels as figures (\ref{fig:meanbar}) and (\ref{fig:sigmabar}) indicate. Hardening the spectrum has the simultaneous effect of raising the amplitude of the resonance modes and decreasing the amplitudes of modes which cause suppression 
so we expect a strong dependence upon $\alpha$ up to the point where the combination of large amplitude and hardness of the turbulence power spectrum means the antineutrinos reach depolarization. 
Beyond that point, the dependence upon amplitudes and power spectral indices is lost. From figures (\ref{fig:meanbar}) and (\ref{fig:sigmabar}) it appears 
a two-flavor depolarization is approached in $\bar{P}_{12}$ for $E=5\;{\rm MeV}$ and $\alpha \lesssim 2$ and in $\bar{P}_{13}$ for $E=45\;{\rm MeV}$ and $\alpha \sim 4/3$ 
only when $C_{\star} =0.5$. 

Except in these cases of very large amplitudes and hard spectra, inspection reveals the distributions of the probabilities clearly posses two components. 
For example, the distributions for $\bar{P}_{12}$ when $E=45\;{\rm MeV}$ and $C_{\star} =0.5$ are mixtures of sharp, zero-peaked distribution and an exponential distribution. 
When we look closely we often find the fraction of the distribution affected by stimulated transitions decreases with $\alpha$ but, simulataneously, 
the effect of the stimulated transitions grows with $\alpha$. This is seen in figure (\ref{fig:P12bar:15MeV}) where we plot the 
frequency distributions of $\bar{P}_{12}$ at the energy of $E=15\;{\rm MeV}$ and $C_{\star}=0.3$. The two mixing distributions are clearly seen in the bottom panel where $\alpha=2$. 
At $\alpha=4/3$ the mixing distributions are a sharp, zero-peaked distribution and an exponential; at $\alpha=2$ this 
has changed to a sharp, zero-peaked distribution and a two-flavor depolarized distribution. These distributions can be compared with those of $P_{12}$ in figure (\ref{fig:P12:15MeV}) for the same energy. 
The mixing channel $\bar{P}_{13}$ is also, generally, a mixture of exponential and narrow distribution which peaks at zero. 
The two cases shown which do not match this pattern are for the $E=15\;{\rm MeV}$ and $E=45\;{\rm MeV}$ antineutrinos at $C_{\star}=0.5$ and $\alpha \lesssim 5/3$ where the distribution is very close to uniform. 

Finally, the one case where distorted phase effects occur are in $\bar{P}_{12}$ when $E=5\;{\rm MeV}$ and $E=15\;{\rm MeV}$. 
This should be expected because in figure (\ref{fig:Pevolution}) we see that the changes in $P_{12}$ and $\bar{P}_{12}$ are occurring 
when the density is between the H and L resonances for the $E=5\;{\rm MeV}$ and $E=15\;{\rm MeV}$ neutrinos. 


\begin{figure}
\includegraphics[clip,width=\linewidth]{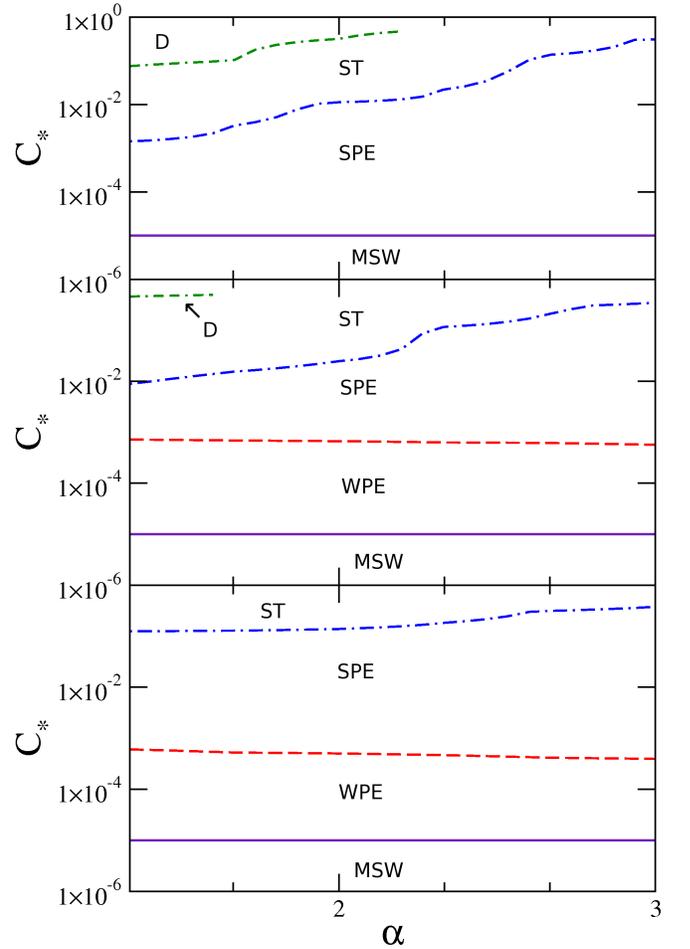}
\caption{The regions of the of the $C_{\star}-\alpha$ plane where the different turbulence effects occur in the H resonant channel $P_{23}$ for different energies. 
In the top panel $E=5\;{\rm MeV}$, in the center $E=15\;{\rm MeV}$, and in the bottom panel $E=45\;{\rm MeV}$ and in each the acronyms stand for 
depolarization (D), stimulated transitions (ST), saturated phase effects (SPE), weak phase effects (WPE), and the
region where only MSW effects occur, i.e.\ no turbulence, is labeled as MSW.\label{fig:regions}}
\end{figure}

\section{Summary and Conclusions}
In this paper we have investigated the effect of modifying the turbulence power spectrum inserted into a supernova density profile upon the neutrinos and antineutrinos for 
three different representative neutrino energies. We have seen the turbulence alters the transitions probabilities of the neutrinos via two effects: the direct stimulation of transitions between the states via 
parametric resonances, and the more subtle effect of changing the phase between semi-adiabatic resonances and/or discontinuities if they are present. 
The two effects depend upon the turbulence power spectrum in different fashions and whether a dependence upon the spectral index is present or not for neutrinos of a given energy 
depends upon the progenitor structure, the postbounce epoch, the turbulence amplitude and the neutrino energy. 

The two most important factors that determine the extent to which turbulence affects the neutrinos is the location of the turbulence in relation to the neutrino's 
resonance densities and the turbulence amplitude. Turbulence effects are largest when the turbulence is located in the profile in the vicinity of the neutrino resonances (both L and H) 
because both effects depend upon the mixing matrix having both large diagonal \emph{and} off-diagonal entries in the region where the turbulence is located and the MSW resonances 
are the locations where these entries are equal in magnitude. In this paper we have used a density profile at a fixed snapshot time and considered three different neutrino 
energies in order to demonstrate this dependence between the turbulence effects and MSW densities. Fixing the neutrino energy and changing the snapshot time would produce similar results. 
In figure (\ref{fig:regions}) we show the regions of $C_{\star}$ and $\alpha$ where we find the various turbulence effects in the mixing between $\nu_2$ and $\nu_3$. 
To construct the figure we have extracted various contours of $\sigma_{23}$ from figure (\ref{fig:sigma}). 
For each energy we find no effect from turbulence when $C_{\star} \leq 10^{-5}$ so have set this as a lower limit in each case. 
For the $E=45\;{\rm MeV}$ neutrinos the boundary between the weak and saturated phase effects is taken to be where $\sigma_{23}=0.1$, the boundary between 
saturated phase effects and stimulated transitions as where $\sigma_{23}=0.2.$. 
For the $E=15\;{\rm MeV}$ neutrinos the boundary between the weak and saturated phase effects is taken to be where $\sigma_{23}=0.3$, 
and this same value of $\sigma_{23}$ forms the boundary between saturated phase effects and stimulated transitions. For the boundary between 
the depolarization region and the stimulated transitions we use $\sigma_{23}=0.25$. Finally, for the $E=5\;{\rm MeV}$ neutrinos 
we do not find a boundary between saturated and weak phase effects, the boundary between saturated phase effects and stimulated transitions is 
where $\sigma_{23}=0.1$ and depolarization is taken to occur when $\sigma_{23} \geq 0.2$. We caution the reader that these boundaries are 
somewhat fuzzy in the sense that on the boundaries the distributions are often mixtures and furthermore these values of $\sigma_{23}$ have no meaning in themselves. 
Let us use this figure to summarize what we have found. 

If we optimize the location of the turbulence by careful selection of the profile, we find the stimulated transition effects appear in every 
mixing channel - even antineutrinos - when the turbulence amplitude exceeds $C_{\star} \gtrsim 1\%$. As one expects for the epoch chosen, the turbulence effects 
are  strongest in the H resonant channel followed by the mixing in $\nu_{1}$-$\nu_{2}$ and $\bar{\nu}_{1}$-$\bar{\nu}_{2}$. 
The sensitivity of the H resonant channel at this epoch is largely the same for all three energies.
The extent to which $\nu_{1}$-$\nu_{2}$ and $\bar{\nu}_{1}$-$\bar{\nu}_{2}$ are affected at this epoch does depend upon the neutrino energy because of the difference of the relation between the turbulent densities 
and the L resonance for different energies. Thus, at this epoch, we see lower energies affected by turbulence to a greater degree than higher energies due to the greater coupling between the states $\nu_1$ and $\nu_2$. 
But even though stimulated transition effects may be present, the neutrinos do not always exhibit a dependence upon the turbulence power spectral index because 
stimulated transitions possess a strong limit of depolarization and a weak limit. In the strong limit no $\alpha$ dependence is found because the distributions of the transition probabilities are depolarized, either two or three flavor; in the weak limit the distributions are found to be exponential and are sensitive to $\alpha$. The boundary between the two regimes shown in figure (\ref{fig:regions})
depends upon $\alpha$ and the turbulence amplitude $C_{\star}$ with a greater proportion of depolarization at fixed $C_{\star}$ when the power spectrum is hardened. 

Stimulated transitions occur when the product of turbulence amplitude and diagonal/off-diagonal mixing matrix elements is large. 
When this product is not above threshold the distorted phase effect of turbulence can become apparent. The boundary between stimulated transitions and phase effects is the dot-dashed line in figure (\ref{fig:regions}). 
In fact, if we optimize the location of the turbulence then the sensitivity to $C_{\star}$ via the distorted phase effect can be extreme with turbulence amplitudes as small as 
$C_{\star} \sim 10^{-4}$ causing an effect. Note this amplitude is comparable with the amplitude of the density fluctuations in the 
progenitor \cite{2011ApJ...733...78A,2015arXiv150302199C} and furthermore the boundary between weak phase effects and the MSW only region in figure (\ref{fig:regions}) varies 
with the epoch: at earlier epochs the boundary lies at much higher values of $C_{\star}$ because the turbulence is far from the H resonant region. 
Like stimulated transitions, the distorted phase effect has a strong and weak limit: in the strong 
limit the phase difference between discontinuities is distributed uniformly leading to arcsine distributions for the transition probabilities. When this occurs 
the parameters describing the arcsine distribution are determined by the jumps in the transition probabilities at the discontinuities and not the turbulence between them. For this reason the 
neutrinos are not affected by changes in the power spectral index. In the weak limit of the distorted phase effect the sensitivity 
to the power spectrum re-emerges due to the dominance of the long wavelength modes. Counter-intuitively, this sensitivity to the longest wavelengths means harder 
spectra have \emph{less} of an effect than soft spectra for a given $C_{\star}$. This difference in the dependence upon $\alpha$ explains the downward slope 
of the boundary between strong and weak distorted phase effects in figure (\ref{fig:regions}).

While the dependence of turbulence effects upon neutrino energy, turbulence amplitude, spectral index etc.~ can be complicated, it is possible to use our results 
to piece together the expected evolution of the turbulence effects that neutrinos of a given energy will experience as a function of time. 
For neutrinos of a given energy there will be no sensitivity to the turbulence during the early phase of the burst signal because the turbulence is at 
densities far from the MSW resonances. This requirement that the turbulence be close to the resonance density is the reason there were no turbulence 
effects seen by Reid, Adams and Seunarine \cite{Reid} when they put turbulence into the post-shock region in profiles appropriate for the accretion epoch.
As the forward shock and turbulence move out into the star turbulence effects will start to appear. 
Initially these are due to distorted phase effects in the H resonance channel and, remarkably, the greater the spectral index $\alpha$ the greater the sensitivity to the turbulence of a 
given amplitude. As time progresses the distorted phase effects for a given neutrino energy will saturate to a limit where the sensitivity to $\alpha$ and $C_{\star}$ is lost. 
As time progresses further stimulated transition effects will begin to appear if the amplitude is greater than $C_{\star} \gtrsim 1\%$. 
Stimulated transitions appear first in the H resonance channel and again, initially, exhibit a sensitivity to $\alpha$ but now we find that smaller values of $\alpha$ lead to 
greater turbulence effects for a given $C_{\star}$. As time progresses further still, yet again that dependence upon $\alpha$ and $C_{\star}$ may disappear 
if the turbulence amplitude is sufficiently great to cause depolarization. If that occurs, turbulence effects will begin to appear in the mixing between other 
states most prominently between $\nu_{1}$-$\nu_{2}$ \emph{and} in the antineutrinos 
in the $\bar{\nu}_{1}$-$\bar{\nu}_{2}$ channel. The mixing in $\nu_{1}$-$\nu_{2}$ and $\bar{\nu}_{1}$-$\bar{\nu}_{2}$ follows the same sequence as the mixing in the H resonant channel i.e.~ 
it starts off as weak distorted phase effects sensitive to $\alpha$ and $C_{\star}$ that then saturates before stimulated transitions appear if $C_{\star}$ is sufficiently large. 
If stimulated transitions do start to affect the $\nu_{1}$-$\nu_{2}$ and $\bar{\nu}_{1}$-$\bar{\nu}_{2}$ evolution while mixing in the H resonant channel is still occurring then it is possible 
in a normal hierarchy to transition to three flavor depolarization if $\alpha$ is sufficiently small and $C_{\star}$ sufficiently large. 
Finally, as the turbulent region sweeps further out into the star, the turbulence effects will decrease in the H resonance channel and concentrate in the 
mixing between $\nu_{1}$-$\nu_{2}$ and $\bar{\nu}_{1}$-$\bar{\nu}_{2}$ which will then themselves eventually fade as the turbulence reaches the very outer edges of the star. 
The extent to which the turbulence effects at these late times are visible will depend upon the exact shape of the progenitor profile because the rapidly fading neutrino luminosity 
will make the statistics of detection very poor. This expected sequence of turbulence events allows us to answer our original question of whether the neutrinos exhibit sensitivity to the turbulence power spectrum. 
We conclude it indeed appears, in principal, there is sensitivity to the power spectral index in the signal from a Galactic supernova and further analyses 
along the lines of Borriello \emph{et al.} \cite{2013arXiv1310.7488B} but for 3D simulations would be very welcome. 


\acknowledgments
This work was supported by DOE grant DE-SC0006417, the Topical Collaboration in Nuclear Science ``Neutrinos and Nucleosynthesis in Hot and Dense Matter'', DOE grant number DE-SC0004786, 
and an Undergraduate Research Grant from NC State University. 


\end{document}